\documentclass[aps,prb,twocolumn,floats,epsfig,showpacs]{revtex4-1}

\usepackage[apple mac]{inputenc}
\usepackage{fourier}
\usepackage{textcomp}

\usepackage{bbm}
\usepackage{amssymb}
\usepackage{ifpdf}
\usepackage{color}
\usepackage{graphicx}
\usepackage{amsmath}
\usepackage{amsfonts}
\usepackage{dcolumn}
\usepackage{bm}
\usepackage{xcolor}

\newcommand{\ket}[1]{\ensuremath{| #1 \rangle}}
\newcommand{\bra}[1]{\ensuremath{\langle #1 |}}

\begin{document}

\title{Rydberg-Atom Quantum Simulation and Chern Number Characterization \\
of a Topological Mott Insulator}

\author{A. Dauphin$^{1,2}$, M. M\"uller$^{1}$, and M. A. Martin-Delgado$^{1}$}
\affiliation{$^1$Departamento de Fisica Teorica I, Universidad Complutense, 28040 Madrid, Spain \\
$^2$Center for Nonlinear Phenomena and Complex Systems, Universit\'{e} Libre de Bruxelles (U.L.B.), Code Postal 231, Campus Plaine, B-1050 Brussels, Belgium}

\begin{abstract}
In this work we consider a system of spinless fermions with nearest and next-to-nearest neighbor repulsive Hubbard interactions on a honeycomb lattice, and propose and analyze a realistic scheme for analog quantum simulation of this model with cold atoms in a two-dimensional hexagonal optical lattice. To this end, we first derive the zero-temperature phase diagram of the interacting model within a mean-field theory treatment. We show that besides a semi-metallic and a charge-density-wave ordered phase, the system exhibits a quantum anomalous Hall phase, which is generated dynamically, i.e.~purely as a result of the repulsive fermionic interactions and in the absence of any external gauge fields. We establish the topological nature of this dynamically created Mott insulating phase by the numerical calculation of a Chern number, and we study the possibility of coexistence of this phase with any of the other phases characterized by local order parameters. Based on the knowledge of the mean-field phase diagram, we then discuss in detail how the interacting Hamiltonian can be engineered effectively by state-of-the-art experimental techniques for laser-dressing of cold fermionic ground-state atoms with electronically excited Rydberg states that exhibit strong dipolar interactions.
\end{abstract}


\maketitle



\section{\label{sec:introduction}Introduction}

Quantum systems and materials with topological properties and quantum order are currently the focus of intense 
theoretical and experimental research: On the one hand, they introduce a new paradigm in condensed matter physics
challenging the standard model of the Landau-Ginzburg theory of phase transitions based on the standard symmetry breaking mechanism. New topological quantum systems exhibit non-local order parameters and require a different framework than the Landau-Ginzburg theory for their understanding and classification\cite{wen-book}. On the other hand, topological quantum systems offer new fascinating potential applications like topological quantum information processing\cite{kitaev-annalsphys-303-2,nayak-rmp-80-1083}.

These developments have not gone unnoticed by another recently emerging field: quantum simulations with cold atoms in optical lattices. 
Originally conceived as platforms to unveil the complicated physics of strongly correlated quantum systems in condensed matter \cite{PhysRevLett.81.3108,greiner-nature-415-39}, its range of applications has rapidly increased \cite{jaksch-ann-phys-315-52,bloch-rmp-2008,lewenstein-adv_amo_phys-56-135,dalibard-rmp-83-1523,lewenstein-book}. The topic has been boosted by enormous progress in the field of cold quantum gases and quantum simulation, where these systems offer unique possibilities to realize and study in a controllable way the physics of some of these models over a wide range of parameters. 

A particular fascinating class of new topological quantum systems are topological insulators (TIs) that provide a new paradigm of how topological properties can be even realized in fermionic free systems under appropriate conditions \cite{hasan-rmp-82-3045,qi-physics-today-2010,qi-rmp-83-1057}. New time reversal TIs were theoretically proposed in two-dimensional hexagonal lattices \cite{haldane-prl-61-2015,kane-prl-95-146802,bernevig-prl-96-106802} and soon extended to three-dimensional arrays 
\cite{fu-prb-76-045302,fu-prl-98-106803,moore-prb-75-121306,qi-prb-78-195424,roy-prb-79-195321,qi-science-323-1184,rosenberg-prb-82-035105}. Remarkably, one-dimensional versions of TIs can be also explicitly realized by means of lattice gauge theory inspired models\cite{creutz-prl-83-2636,bermudez-prl-102-135702,arxiv_viyuela-2012}. In a series of ground-breaking experiments, the existence of this new quantum phases of matter has been unambiguously  confirmed thereby establishing the topic of TIs as a solid and consolidated research ground \cite{koenig-science-318-766,hsieh-nature-452-970,roth-science-325-294,day-physics-today-2008,chen-science-325-178,xia-natphys-5-398}.

There are two main avenues to realize TI phases:

i/ The original approach to simulate TI systems with cold atoms in optical lattices stems from the idea of synthesizing gauge magnetic fields in an effective way \cite{jaksch-njp-5-56,mueller-pra-70-041603,sorensen-prl-94-086803,spielman-pra-79-063613,gerbier-njp-12-033007}, which has recently been demonstrated experimentally\cite{lin-nature-462-628,aidelsburger-prl-107-255301}. The reasoning behind these quantum engineering ideas is that background fixed classical gauge fields play a fundamental role in the theoretical construction of topological insulating systems. As an example of the broad range of possibilities of this new line of quantum simulation compared to the study of standard condensed matter systems in nature, it is worth mentioning the possibility of simulating non-Abelian gauge fields in several different cases \cite{osterloh-prl-95-010403,ruseckas-prl-95-010404,goldman-prl-103-035301,goldman-pra-79-023624,bermudez-njp-12-033007,satija-prl-97-216401,satija-pra-77-043410,Mazza-njp-14-015007,arxiv_hauke-1205.1398-2012}, in particular an anomalous relativistic non-Abelian quantum Hall effect \cite{goldman-pra-79-023624}. These synthetic gauge field implementations have paved the way to the quantum simulation of TIs with optical lattices, ranging from proposals for one-dimensional systems\cite{Mazza-njp-14-015007,lang-prl-108-220401,arxiv_kraus-1109.5983-2012,mei-pra-85-013638}, two-dimensional lattices\cite{zhu-prl-97-240401,umucalilar-prl-100-070402,stanescu-pra-79-053639,goldman-prl-105-255302,stanescu-pra-82-013608} even up to three-dimensional implementations\cite{bermudez-prl-105-190404,beri-prl-107-145301} involving realizations of lattice gauge theory effects.

ii/ Alternatively, there exists the less explored possibility that \textit{fermionic interactions} lead to the dynamical creation of a TI phase: Understanding the role of fermionic interactions in TIs is attracting a lot of attention since the classification of non-interacting TIs and superconductors has been achieved in the form of the celebrated "Periodic Table"\cite{ptable1,ptable2}. This classification is exhaustive for quadratic fermionic Hamiltonians and depends on the dimensionality of the lattice fermions and the discrete anti-unitary symmetries which are broken or not by the dynamics, specifically, time-reversal and particle-hole symmetry as well as chirality. Interestingly enough, with the pioneering work by Raghu \textit{et al.}\cite{Raghu-prl-100-156401} and subsequent studies\cite{Weeks-prb-81-085105,zhang-prb-79-245331,sun-prl-103-046811,wen-prb-82-075125,varney-prb-84-241105,PhysRevLett.107.106402} it has become clear that interactions in purely fermionic Hamiltonians may not only change the topological class of a TI into a different (possibly trivial) one. Instead, the interactions may actually become a \textit{resource} to produce TI phases. The key point of this new mechanism is that there is no need to resort to background-fixed gauge fields coupled to fermion degrees of freedom, but that it is the very presence of many-body interaction effects among fermions hopping in particular lattice geometries that produces a new phase called a Topological Mott Insulator (TMI). 

In this work we are interested in the latter interacting scenario ii/ as a way to produce TI phases whose non-trivial topological properties arise purely as a consequence of the fermionic interactions, as opposed to the scenario i/ of gauge-field induced TIs. We will consider a model of interacting spinless fermions on a hexagonal lattice, which we specify below, where this mechanism may result in the formation of a TMI phase, generated dynamically by repulsive Hubbard-type interactions and in the absence of any (synthetic) external gauge fields. Besides a theoretical analysis of the model, we will propose how this new physics could be observed in an experiment by implementing the interacting fermion model in a quantum simulator using cold Rydberg atoms in optical lattices. On the one hand, such quantum simulation is important for benchmarking theoretical predictions with experimental observations, in particular as the problem of interacting lattice fermions in two or higher spatial dimensions is notoriously hard to solve. On the other hand, as we will show below, the flexibility to tune the parameters of the model of interest over a wide range of parameters in a quantum simulation offers the possibility to explore in the laboratory the physics in regimes, which are not naturally realized or not readily accessible in known materials. 

In the context of quantum simulation, we will see that as an alternative to engineer complex next-to-nearest neighbor hopping dynamics as required in the Haldane model\cite{haldane-prl-61-2015}, which supports a quantum anomalous Hall (QAH) phase that has so far not been observed in any material, introducing long-range interactions between the fermions might become an ally, as long as these turn out to favor the formation of a stable QAH phase and there is no need to implement synthetic gauge fields. We anticipate at this stage that the fact that both the Haldane model and the interacting model considered by Raghu \textit{et al.}\cite{Raghu-prl-100-156401} and also by us in this work are defined on the same type of hexagonal lattice is relevant for the following reason: the next-to-nearest neighbor fermionic interactions,  when treated within a mean-field theory (MFT) approach, give rise to an order parameter that resembles the single-particle next-to-nearest neighbor hopping in the Haldane model. This connection is the physical key mechanism underlying the dynamically created, i.e.~interaction-induced TMI phase described above.

Finally, we remark that the hexagonal lattice geometry considered in this work and our results are also related to the very rich physics in graphene\cite{novoselov-nature-438-197,peres-prb-73-125411,castro-rmp-81-109}, where effective gauge fields can be induced by means of disclinations or strain fields\cite{vozmediano-physrep-496-109}, as well as irradiation with light\cite{kitagawa-prb-84-235108,kitagawa-prb-82-235114}.

\subsection{\label{subsec:structure_of_the_paper} Structure of the paper}

In this work we 
\begin{enumerate}
\item
identify the relevant order parameters and determine the complete zero-temperature phase diagram within a MFT treatment, and compare our findings to related, previous work on this model\cite{Raghu-prl-100-156401,Weeks-prb-81-085105},
\item
study the possibility of coexistence of the encountered topologically non-trivial Mott insulating phase with any of the other phases with local order parameters,
\item
underpin the character of the topologically trivial and non-trivial phases by an explicit numerical calculation of Chern numbers\cite{thouless-prl-49-405,niu-prb-31-3372,kohmoto-ann-phys-160-343,hatsugai-prb-48-11851,sheng-prl-97-036808,bib:fukui,fukui-prb-75-121403}, which we compare with the behavior of local order parameters,
\item
propose and analyze a realistic scheme for an analog quantum simulation of the model with cold fermionic Rydberg atoms in an optical lattice. 
\end{enumerate}

This paper is organized as follows: in Sect.~\ref{sec:model} we introduce
the model Hamiltonian describing spinless fermions on a hexagonal lattice
subject to Hubbard interactions. Then, we go on to review the basic
properties of the free hopping Hamiltonian part, in particular the
momentum representation of quantum states in the Brillouin zone
that constitute the basis for the rest of the analysis.
In Sect.~\ref{sec:meanfield_and_phasediagram}, we perform 
a MFT analysis for dealing with the Hubbard
interaction terms, which allows us to identify and characterize the relevant phases in the various parameter regimes. In Sect.~\ref{sec:qs_Rydberg_atoms} we show how the model Hamiltonian of interacting fermions can be engineered and arises as an effective description of the dynamics of fermionic ground state atoms laser-coupled to Rydberg states, which exhibit strong and long-range interactions. Finally, Sect.\ref{sec:conclusions_and_outlook} is devoted to conclusions and an outlook.

Two appendices contain the detailed and basic
explanations of the constructions used throughout the text.
Specifically, Appendix \ref{app:Chern_numbers} presents the details on the
calculation of the Chern numbers characterizing the topological Mott
insulating phase deduced from the MFT analysis. Appendix \ref{app:brillouin_zone} summarizes
details on various parametrizations of the Brillouin zone and domain of integration.

\section{\label{sec:model}The model}

Let us consider spinless fermions at sites $j\in \Lambda$ of a hexagonal lattice in two dimensions. The associated fermionic creation $c_j^\dag$ and annihilation operators $c_j$ satisfy the canonical anti-commutation relations $\{c_i,c_j^\dag \}=\delta_{i,j}$. The model Hamiltonian $H$ we are interested in is an extension of the free fermion hopping Hamiltonian $H_t$ between nearest-neighbor sites of the lattice by including Hubbard-type nearest-neighbor $H_{\rm NN}$ and next-to-nearest neighbor $H_{\rm NNN}$ interactions: 
\begin{equation}
\begin{split}
H&:= H_t + H_{\rm NN} + H_{\rm NNN} \\
&:=-t\sum_{ \langle i,j  \rangle  }(c^\dagger_ic_j+h.c.)+V_1\sum_{ \langle i,j  \rangle  }n_i n_j+V_2\sum_{ \langle \langle i,j  \rangle \rangle }n_i n_j \text{,}
\end{split}
\label{hamiltonian}
\end{equation}
where $ \langle i,j  \rangle $ denotes the summation over pairs of nearest-neighbor, and $ \langle \langle i,j  \rangle \rangle $ over the next-to-nearest neighbor sites, and $n_i = c_i^\dagger c_i$ denotes the number operator on lattice site $i$. The real-valued hopping amplitude is $t$, while the interacting couplings are repulsive $V_1,V_2>0$.

This model has appeared in different contexts, such as the study of antiferromagnetic phases
using quantum Monte Carlo simulations \cite{sorella_tosatti_92}, and as a candidate for a quantum spin liquid \cite{hermele_07,meng-nature-464-847}. Here, our focus lies on a quantum simulation of the model with Rydberg atoms in optical lattices. To connect with a physical implementation in a cold atoms setup, we will first study the model theoretically which allows us to establish its phase diagram and identify the relevant order parameters and observables to be measured in an experiment.

\begin{figure}[t]
\begin{center}
	\includegraphics[width=0.9\columnwidth]{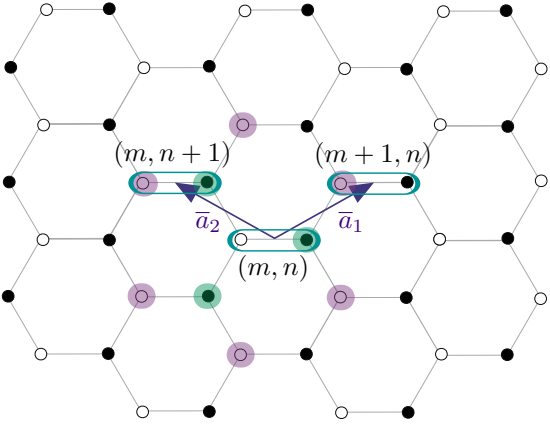}\hspace{0.8cm}
	\caption{Hexagonal lattice $\Lambda$ with sites $j$ decomposed in two-site basis cells $j=(m,n)$ connected by basis vectors $\mathbf{a}_1$ and $\mathbf{a}_2$. A fermion at the $\phi$-type site $(m,n)$ possesses three nearest-neighbor sites (shaded black filled circles) and six next-to-nearest neighbor sites (shaded white open circles).}
\label{fig:honeycomb}
\end{center}
\end{figure} 

Before dealing with the full Hamiltonian (\ref{hamiltonian}) we start first by reviewing the properties of the free part of the Hamiltonian $H_t$, in order to establish the basic notations and operators that we shall use below when treating the interaction terms in Sect.\ref{sec:meanfield_and_phasediagram}. The bipartite nature of the hexagonal lattice makes it convenient to rewrite the Hamiltonian $H_t$ in terms of two-site basis cells\cite{bib:bena} (see Fig.~\ref{fig:honeycomb}), labeled by an index pair $(m,n)$, each of which contains one site of the $\phi$-sublattice (open white circles) and one of the $\psi$-sublattice (filled black circles): 
\begin{equation}
H_{t}=-\sum_{m,n} t (c^\dagger_{\psi mn} c_{\phi mn}+c^\dagger_{\psi mn+1} c_{\phi mn}+c^\dagger_{\psi m-1n} c_{\phi mn})+h.c. \; \text{.}
\end{equation}
Correct counting of terms, thus avoiding possible double counting errors, is essential, in particular to correctly reproduce the quantitative details of the mean-field phase diagram of (\ref{hamiltonian}), such as the values of the critical couplings where phase transitions take place. We will comment on some controversy on this issue in the recent literature\cite{Raghu-prl-100-156401,Weeks-prb-81-085105} in more detail below. At this point we would like to note that the sum over $n,m$ of the hopping processes of a fermion at a site $\phi$-type lattice site ($m,n$) to one of its three neighboring $\psi$-type sites already covers all possible hopping processes along all links connecting neighboring sites of the hexagonal lattice. 

In addition to the tight-binding term, we consider a staggering potential, describing a chemical potential difference for fermions residing at $\phi$- and $\psi$-sites, which reads
\begin{equation}
H_{\beta}=\beta \sum_{m,n}  (c^\dagger_{\phi mn} c_{\phi mn}  - c^\dagger_{\psi mn} c_{\psi mn} )  \; \text{.}
\end{equation}
To determine the energy spectrum of the tight-binding Hamiltonian in combination with the staggering potential, we introduce Fourier-transformed fermionic operators, 
\begin{equation}
c_{\alpha mn}  = \frac{1}{\sqrt{N_\Lambda}} \sum_{\mathbf{k} \in \text{BZ}} \exp(i \mathbf{k}\, \cdot \mathbf{r}_{nm} ) c_\alpha (\mathbf{k}), 
\end{equation}
with $\alpha \in \{ \phi, \psi \}$, to rewrite the Hamiltonian in momentum space as
\begin{equation}
H_t + H_\beta = \sum_{\mathbf{k}} 
  \hat{\Psi}^\dagger(\mathbf{k})
 \begin{pmatrix}
 \beta & -tA_\mathbf{k}^* \\-tA_\mathbf{k} & -\beta
 \end{pmatrix}
  \hat{\Psi}(\mathbf{k}).
 \label{eq:Hamiltonian_hopping_and_staggering}
\end{equation}
Here, $ \hat{\Psi}^\dagger(\mathbf{k}) = \left( 
c^\dagger_\phi(\mathbf{k}), \, c^\dagger_\psi(\mathbf{k}) \right)$, 
and $\mathbf{r}_{mn} = n \mathbf{a}_1 + m \mathbf{a}_2$ are the real-space vectors of the lattice, with 
\begin{equation}
\mathbf{a}_1 = (3/2,\,\sqrt{3}/2), \qquad \mathbf{a}_2 = (-3/2,\,\sqrt{3}/2)
\end{equation}
denoting the basis vectors (for a unit length lattice spacing), which span the whole lattice in the two-site basis parametrization (see Fig.~\ref{fig:honeycomb}). The total number of two-site basis cells is $N_\Lambda$, which is half of the total number of lattice sites, $N_\Lambda = N/2$. The function 
\begin{equation}
A_\mathbf{k} = 1+\exp( i\mathbf{k}\,\cdot \mathbf{a}_1) + \exp (-i\mathbf{k}\,\cdot \mathbf{a}_2)
\end{equation}
contains the information about the structure and symmetry of the hexagonal lattice. The reciprocal lattice vectors, symmetry properties and various equivalent parametrizations of the Brillouin zone, either in a discretized or in a continuous form, are provided in Appendix \ref{app:brillouin_zone}. 
 
Diagonalization of Hamiltonian (\ref{eq:Hamiltonian_hopping_and_staggering}) readily yields the two-band energy spectrum 
\begin{align}
&E_\pm (\mathbf{k}) = \pm \sqrt{\beta^2+t^2|A_\mathbf{k}|^2} \\ 
&= \pm \sqrt{\beta^2+t^2 \left( 3+2\cos( \sqrt{3}k_y) +4\cos (\sqrt{3}k_y/2) \cos(3 k_x/2) \right) } \nonumber
\end{align}
Whereas a non-vanishing staggering potential $\beta$ opens an energy gap between the two bands, for $\beta = 0$ the gap closes and the spectrum exhibits the well-known band structure\cite{Wallace-PhysRev-71-622,Semenoff-prl-53-2449,Redlich-prl-52-18,DiVincenzo-prb-B-29-1685,Gonzalez-nucl-phys-b-406-771} with a linear dispersion relation in the vicinity of the Dirac cones (see Fig.~\ref{fig:diraccone} in Appendix \ref{app:brillouin_zone}). Our motivation to consider the staggering potential $H_\beta$ at this stage is because interacting terms $H_{\rm NN}$ and $H_{\rm NNN}$, when treated within a MFT approach in the following section, will produce in momentum space a very similar structure. In fact, we anticipate that -- similar to the effect of an arbitrarily weak staggering potential -- repulsive nearest- and next-to nearest-neighbor interactions above a critical strength can induce openings of an energy gap, associated to first- and second-order quantum phase transitions. 

\section{\label{sec:meanfield_and_phasediagram} Mean-field analysis and phase diagram}

\subsection{Mean-field treatment for nearest-neighbour interactions}
As a first step, and for clarity of the presentation, let us for the moment consider the mean-field treatment of the Hamiltonian (\ref{hamiltonian}) only including nearest-neighbor interactions, 
\begin{equation}
H_1 := H_t + H_{\rm NN}.
\end{equation}
This analysis presents some of the basic ingredients that will later be used in the study of the complete Hamiltonian, and introduces the charge-density-wave (CDW) order parameter as one of the basic order parameters of the system.

We perform the mean-field approximation of the interacting part, which is quartic in fermionic operators, according to the Wick theorem\cite{bruus-flensberg-book}:
\begin{align}
\label{eq:meanfieldapprox}
n_i n_j \simeq & \langle c_i^\dagger c_i \rangle c_j^\dagger c_j + \langle c_j^\dagger c_j \rangle c_i^\dagger c_i -  
\langle c_i^\dagger c_i \rangle \langle c_j^\dagger c_j \rangle \nonumber \\
&  - \left( \langle c_i^\dagger c_j \rangle c_j^\dagger c_i + \langle c_j^\dagger c_i \rangle c_i^\dagger c_j -  
\langle c_i^\dagger c_j \rangle \langle c_j^\dagger c_i \rangle \right).
\end{align}
By introducing the self-consistently defined mean values $\overline{n}_i := \langle c_i^\dagger c_i \rangle $ and $\xi_{ij} := \langle c_i^\dagger c_j \rangle$ the Hamiltonian $H_1$ in mean-field approximation takes the form
\begin{align}
H_{1{\rm MF}} = & - \sum_{ \langle i,j  \rangle} \left[ (t+V_1 \xi_{ij}) c^\dagger_j c_i  + h.c. - V_1 |\xi_{ij}|^2 \right] \nonumber \\
& + V_1 \sum_{ \langle i,j  \rangle} \left( \overline{n}_i  n_j +\overline{n}_j n_i  - \overline{n}_i\overline{n}_j  \right).
\end{align}
One sees that the presence of $H_{\rm  NN}$ leads in mean-field approximation, besides the repulsive density-density-type interaction terms,  to a renormalization of the bare hopping amplitude $t$,
\begin{equation}
t \rightarrow t'  = t + \xi V_1.
\label{eq:t_to_t'_renormalization}
\end{equation}
Similar to the rewriting of $H_t$ above, we now reformulate $H_{1{\rm MF}}$ in the two-site basis, and in addition assume that the expectation values of the fermion density on $\phi$- and on $\psi$-sites of the lattice, $\overline{n}_\phi$ and $\overline{n}_\phi$, are translationally invariant, and that $\xi := \langle c_\phi^\dagger c_\psi \rangle$ is real-valued and rotationally invariant. We thus obtain the mean-field Hamiltonian in momentum space
\begin{align}
H_{1{\rm MF}} = \sum_{\mathbf{k}} 
  \hat{\Psi}^\dagger(\mathbf{k}) \mathcal{H}_{1{\rm MF}}(\mathbf{k}) 
  \hat{\Psi}(\mathbf{k})- 3 N_\Lambda V_1 (\overline{n}_\phi \overline{n}_\psi - \xi^2)
\label{eq:mean_field_Ham1_momentum_space}
\end{align}
with 
\begin{equation}
\mathcal{H}_{1{\rm MF}} (\mathbf{k}) =  
\begin{pmatrix}
3 V_1 \overline{n}_\psi & - (t + V_1 \xi) A_\mathbf{k}^* \\ - (t + V_1 \xi) A_\mathbf{k} &  3 V_1 \overline{n}_\phi 
 \end{pmatrix}.
  \label{eq:mean_field_Ham1_momentum_space_matrix}
 \end{equation}
Nearest-neighbor repulsive interactions are known to break the symmetry of $\phi$- and $\psi$-lattice sites and to favor the formation of a CDW\cite{Raghu-prl-100-156401}. We define the CDW order parameter
\begin{equation}
\rho:=\frac{1}{2}(\overline{n}_\phi-\overline{n}_\psi)
\end{equation}
which quantifies the fermion density imbalance at $\phi$ and $\psi$-sites of the two sub-lattices, for a total density of 
\begin{equation}
n: = \frac{1}{2}(\overline{n}_\phi + \overline{n}_\psi)
\end{equation}
in the system. Diagonalization of (\ref{eq:mean_field_Ham1_momentum_space_matrix}) yields the energy-momentum dispersion relation 
\begin{equation}
E_\pm (\mathbf{k}) = 3 V_1n \pm \sqrt{(3V_1 \rho)^2 +  (t + V_1 \xi)^2 |A_\mathbf{k}|^2}.
\end{equation}
At zero temperature, the lower energy band is completely filled, and the free energy (per two-site basis cell) is
\begin{equation}
F/N_\Lambda = 3V_1(\rho^2 + \xi^2) - 3V_1n^2 + \frac{1}{L^2}\int_{\rm BZ} d^2k\,  E_-(\mathbf{k}).
\label{eq:free_energy_H1MF}
\end{equation}
Here, we have converted the discrete sum over momenta into an integral over the first Brillouin zone, with $L^2$ denoting the integration area in momentum space (see Appendix \ref{app:brillouin_zone} for details). 

In which quantum phase the system is, as a function of the coupling parameters $V_1$ and $t$, is determined by the global minimum of the free energy functional. Evaluation of the partial derivatives $\partial F /\partial \xi = 0$ and $\partial F /\partial \rho = 0$ yields the set of coupled self-consistency equations for the CDW order parameter $\rho$ and the renormalization (\ref{eq:t_to_t'_renormalization}) of the nearest-neighbor hopping amplitude $\xi$,
\begin{align}
\label{eq:xi_equation_H1MF}
\xi & = \frac{1}{6L^2}\int_{\rm BZ}  d^2k \frac{(t+\xi V_1)|A_\mathbf{k}|^2}{\sqrt{(3V_1\rho)^2+(t+\xi V_1)^2|A_\mathbf{k}|^2}},\\
\label{eq:rho_equation_H1MF}
\rho & = \frac{3V_1}{2L^2}\int _{\rm BZ} d^2k \frac{\rho}{\sqrt{(3V_1\rho)^2+(t+\xi V_1)^2|A_\mathbf{k}|^2}}.
\end{align}
In the discussion of the phase diagram below, these equations will serve us to study the renormalization of the hopping amplitude and to determine the critical interaction strength $V_{1c}$ (for vanishing next-to-nearest neighbor interactions $V_2$) at which a phase transition from a SM region to a CDW phase takes place.

\subsection{\label{subsec:MFT_H2}Mean-field analysis of the next-to-nearest neighbor interactions}

In this subsection we analyze the effect of the $H_2$-term in a mean-field treatment, defining the auxiliary Hamiltonian 
\begin{equation}
\label{eq:H_2}
H_2 := H_t + H_{\rm NNN}.
\end{equation}
As for the nearest-neighbor interactions, we decouple the quartic interaction terms according to the mean-field approximation (\ref{eq:meanfieldapprox}). 

\begin{figure}[t]
\begin{center}
\includegraphics[scale=0.75]{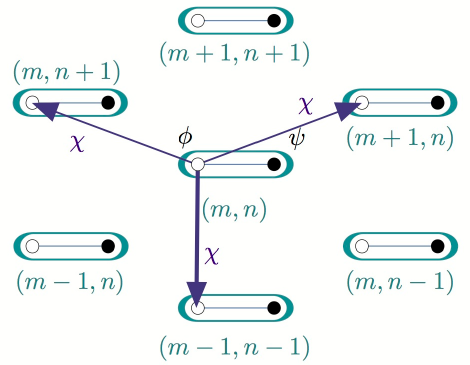}
\newline
\includegraphics[scale=0.8]{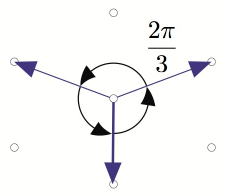}
\includegraphics[scale=0.7]{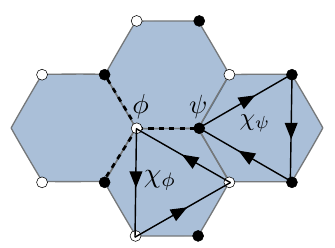}
\caption{The rotational symmetry illustrated in this figure, in combination with translational symmetry, leads to the identification given by Eqs.~(\ref{eq:chi_alpha_1st_equation}) and (\ref{eq:chi_alpha_2nd_equation}).}
\label{fig:chi_illustration}
\end{center}
\end{figure}

The diagonal part (first line of (\ref{eq:meanfieldapprox})) leads to an effective next-to-nearest neighbor density-density interactions of fermions residing at sites of the \textit{same} sublattice $\alpha \in \{ \phi, \psi\}$. Inspection of one representative two-site basis cell $(m,n)$ shows that an $\alpha$-site possesses six next-to-nearest-neighbor $\alpha$-sites.  Each of these six next-to-nearest neighbor interaction terms produces a mean-field contribution $(2 \overline{n}_\alpha n_\alpha - \overline{n}_\alpha^2)$, which, however, is to be counted only with weight $1/2$ to avoid double-counting, when the sum over all two-site basis cells of the lattice is carried out. In consequence, after performing the transformation to momentum space the resulting density-type contribution to the Hamiltonian reads
\begin{equation}
H_{2 \rm{MF}}^{\rm{density}} = 3 V_2 \sum_{\mathbf{k}} \sum_{\alpha} (2 \overline{n}_\alpha n_\alpha - \overline{n}_\alpha^2)
\label{eq:H2MF_density}
\end{equation}
The off-diagonal terms (second line of (\ref{eq:meanfieldapprox})) play a fundamentally different role: They correspond to emerging next-to-nearest neighbor hopping processes, $\chi_{\alpha, m,n, m', n'} c^\dagger_{\alpha m'n'} c_{\alpha mn}$, with the self-consistently defined mean values $\chi_{\alpha, m,n, m', n'} := \langle c^\dagger_{\alpha mn} c_{\alpha m'n'}\rangle$. A mean-field ansatz for the $\chi_\alpha$ mean values on the two sublattices, which (i) is translationally invariant, i.e.~under displacements according integer multiples of the real-space lattice vectors $\mathbf{a}_1$ and $\mathbf{a}_2$, (ii) which still possesses a rotational symmetry under rotations about an angle of $2 \pi/3$, and (iii) is compatible with complex-valued $\chi$-fields, is given by the identification (see Fig.~\ref{fig:chi_illustration}) 
\begin{align}
\label{eq:chi_alpha_1st_equation}
\chi_\alpha & \equiv \chi_{\alpha,m+1,n,m,n}=\chi_{\alpha,m,n+1,m,n}=\chi_{\alpha,m-1,n-1,m,n},\\
\chi_\alpha^* & \equiv \chi_{\alpha,m+1,n+1,m,n}=\chi_{\alpha,m-1,n,m,n}=\chi_{\alpha,m,n-1,m,n}.
\label{eq:chi_alpha_2nd_equation}
\end{align}
It will be convenient to parametrize the two mean fields in terms of their absolute value and phase,
\begin{equation}
\chi_{\alpha} = |\chi_\alpha| \,e^{i \Theta_\alpha}, \qquad \alpha \in \{ \phi, \psi \},
\label{eq:chi_parameters}
\end{equation}
and as we will see below in the study of the complete Hamiltonian and its mean-field phase diagram the $\chi$-fields will be the order parameter associated to the appearance of a topologically non-trivial phase. 

The contribution to the mean-field Hamiltonian $H_{2 \rm MF}$ arising from these off-diagonal terms, after transforming to Fourier space is given by
\begin{equation}
H_{2 \rm{MF}}^{\rm{off-diag}} = - V_2 \sum_{\mathbf{k}} \sum_{\alpha} \left( 2 |\chi_\alpha| f_{\mathbf{k}}(\Theta_\alpha) n_\alpha - 3 |\chi_\alpha|^2 \right)
\label{eq:H2MF_offdiag}
\end{equation}
with the momentum-dependent function
\begin{equation}
f_{\mathbf{k}}( \Theta) := \cos (\mathbf{k}\, \cdot \mathbf{a}_1 + \Theta ) + \cos (\mathbf{k}\, \cdot \mathbf{a}_2 + \Theta )  + \cos (\mathbf{k}\, \cdot (\mathbf{a}_1+\mathbf{a}_2) - \Theta ).
\label{eq:f_k}
\end{equation}
Combining the two contributions (\ref{eq:H2MF_density}) and (\ref{eq:H2MF_offdiag}) with the tight-binding term $H_t$ yields the expression for the complete mean-field Hamiltonian 
\begin{align}
H_{2{\rm MF}} = & \sum_{\mathbf{k}} 
  \hat{\Psi}^\dagger(\mathbf{k}) \mathcal{H}_{2{\rm MF}}(\mathbf{k}) 
  \hat{\Psi}(\mathbf{k})  \nonumber \\
   & \qquad + 3 N_\Lambda V_2 \left( | \chi_\phi |^2 +  | \chi_\psi |^2 - ( \overline{n}^2_\phi + \overline{n}^2_\psi ) \right)
 \label{eq:mean_field_Ham2_momentum_space}
\end{align}
with 
\begin{equation}
\mathcal{H}_{2{\rm MF}}(\mathbf{k}) =
 \begin{pmatrix}
 V_2 (6 \overline{n}_\phi - 2  | \chi_\phi | f_{\mathbf{k}}(\Theta_\phi) ) & - t \, A_\mathbf{k}^* \\ - t \, A_\mathbf{k} &  V_2 (6 \overline{n}_\psi - 2  | \chi_\psi | f_{\mathbf{k}}(\Theta_\psi) ) 
 \end{pmatrix}.
 \end{equation}

\subsection{\label{subsec:complete_MF_Ham}Complete Hamiltonian in the mean-field approximation}

By combining the results of the two previous subsections, we are in the position to formulate the mean-field approximation for the complete Hamiltonian (\ref{hamiltonian}). In momentum space it reads
\begin{align}
H_{\rm MF} & = \sum_{\mathbf{k}} 
  \hat{\Psi}^\dagger(\mathbf{k}) \mathcal{H}_{\rm MF}(\mathbf{k}) 
  \hat{\Psi}(\mathbf{k})  \nonumber \\
   & + 3 N_\Lambda \left( V_1 \left( \xi^2 - \overline{n}_\phi \overline{n}_\psi \right) + V_2  \left( | \chi_\phi |^2 +  | \chi_\psi |^2 - ( \overline{n}^2_\phi + \overline{n}^2_\psi ) \right)\right)  
    \label{eq:complete_mean_field_Ham_momentum_space}
\end{align}
with the matrix elements of $\mathcal{H}_{\rm MF}$ given by
\begin{align}
\mathcal{H}_{{\rm MF}}^{(11)}(\mathbf{k}) = & 3V_1 \overline{n}_\psi + 6 V_2 \overline{n}_\phi - 2 V_2  | \chi_\phi | f_{\mathbf{k}}(\Theta_\phi) \\
\mathcal{H}_{{\rm MF}}^{(22)}(\mathbf{k}) = & 3V_1 \overline{n}_\phi + 6 V_2 \overline{n}_\psi - 2 V_2  | \chi_\psi | f_{\mathbf{k}}(\Theta_\psi) \\
\mathcal{H}_{{\rm MF}}^{(12)} (\mathbf{k})= & \left( \mathcal{H}_{{\rm MF}}^{(21)}(\mathbf{k}) \right)^* = - (t+ \xi V_1)\, A_\mathbf{k}^*
 \end{align}
The Hamiltonian is readily diagonalized, yielding the zero-temperature free energy 
\begin{align}
\label{eq:total_free_energy}
F/N_\Lambda =  & \,\,3V_1 \xi^2 + 3(V_1 - 2V_2) \rho^2  + 3 V_2 \left( | \chi_\phi |^2 +  | \chi_\psi |^2 \right) \\
& - 3 (V_1 + 2V_2) (n^2-n) \nonumber \\
& - V_2 \frac{1}{L^2}\int _{\rm BZ} d^2k \left( | \chi_\phi | f_{\mathbf{k}}(\Theta_\phi)+  | \chi_\psi | f_{\mathbf{k}}(\Theta_\psi) \right) \nonumber \\
& -  \frac{1}{L^2}\int _{\rm BZ} d^2k \left[ (t+ \xi V_1)^2 |A_\mathbf{k}|^2 \right. \nonumber \\
& \,\,\,\left. + \left( 3(V_1 -2V_2) \rho + V_2  \left( | \chi_\phi | f_{\mathbf{k}}(\Theta_\phi) - | \chi_\psi | f_{\mathbf{k}}(\Theta_\psi) \right) \right)^2 \right]^{1/2} \nonumber
\end{align}
where the expressions involving $\overline{n}_\alpha$ have been re-expressed in terms of the CDW order parameter $\rho$ and the total fermion density $n$. Before starting with the calculation of the phase diagram, based on the free energy functional (\ref{eq:total_free_energy}), we add a few remarks:

First, Eq.~(\ref{eq:total_free_energy}) compared to Eq.~(\ref{eq:free_energy_H1MF}) shows that the repulsive next-to-nearest neighbor interactions, which appear as the combination $3(V_1 -2V_2) \rho$ in the integrand and in form of the constant $3(V_1 - 2V_2) \rho^2$ in the free energy functional, effectively reduce the effect of the nearest-neighbor interactions, favoring that the system is in a phase with homogenous fermion density over the lattice, thereby tending to suppress the formation of a phase with CDW order. However, for too strong next-to-nearest neighbor interactions $V_2$, once $(V_1 -2V_2) < 0$, one enters a regime in which in the employed two-lattice-site mean-field approximation the effective interactions appear as if they were attractive
\footnote{The physics of attractive spinless fermions in a honeycomb-lattice is qualitatively different and has been recently studied: see e.g.~D. Poletti, C. Miniatura and B. Gr\'emaud, Euro. Phys. Lett.~{\bf 93}, 37008 (2011); P. Corboz, S. Capponi, A. M. Laeuchli, B. Bauer and R. Orus, Euro. Phys. Lett.~{\bf 98}, 27005 (2012)}. 
In this limit the mean-field description used here obviously cannot be expected to reproduce reliable results. This deficiency can be dealt with by: (i) either working with a spatially enlarged MFT ansatz, which includes not a single two-site cell, but e.g.~a six-site hexagon as the elementary spatial unit\cite{Weeks-prb-81-085105}. Such enlarged, more sophisticated ansatz is able to correctly capture the repulsive character of the next-to-nearest neighbor interactions within the mean-field approximation. (ii) Alternatively, one can stick to the spatially reduced two-site cell ansatz, and set for the described physical reasons the correction due to $V_2$ in the $3(V_1 -2V_2) \rho$ and $3(V_1 -2V_2) \rho^2$ terms in the free energy expression by hand to zero. This corresponds to not including the density-type decoupling channel in the mean field approximation of the quartic $V_2$ term $H_{\rm NNN}$ (i.e.~in suppressing the three terms resulting from the first line of Eq.~(\ref{eq:meanfieldapprox}) in the treatment of Hamiltonian (\ref{eq:H_2}) that lead to the contribution (\ref{eq:H2MF_density})). As our main focus lies on the quantum simulation of the interacting fermion model on a cold atom platform, we follow the second approach (ii), which has also been adopted in previous work by Raghu \textit{et al.} \cite{Raghu-prl-100-156401}.

Second, we remark that in the above derivation so far no assumptions have been made on the absolute values and phases of the order parameters $\chi_\phi$ and $\chi_\psi$ (see Section \ref{subsec:MFT_H2} and Eq.~(\ref{eq:chi_parameters})). To determine the phase diagram in the next section, we will adopt the restricted ansatz
\begin{equation}
\chi_{\phi} =+i |\chi|,\qquad \chi_{\psi} =-i |\chi|,
\label{eq:chi_parameters_reduced}
\end{equation}
with fixed phases $\Theta_\phi = \pi/2$, $\Theta_\psi = -\pi/2$, and equally large absolute values. This parametrization has been adopted previously in the literature\cite{Raghu-prl-100-156401,Weeks-prb-81-085105}. Here we also choose this ansatz, also with the purpose to clarify a controversy about the numerical values of critical points in the phase diagram. In Section \ref{subsec:coexistence}, we will return to the question, whether the spatially limited mean-field ansatz involving only two lattices sites, allows for richer physics, if one does not a priori fix the $\chi_\alpha$ according to Eq.~(\ref{eq:chi_parameters_reduced}), which only involves a single real-valued parameter $|\chi|$. In particular we will discuss whether the more general ansatz (\ref{eq:chi_parameters}) allows for co-existence of a topologically non-trivial phase and a CDW  phase.

Taking into account the described modifications, the free energy (\ref{eq:total_free_energy}) reduces to
\begin{align}
F/N_\Lambda = 3V_1 \xi^2 + 3V_1\rho^2  + 6 V_2 | \chi |^2 - 3 (V_1 + 2V_2) (n^2-n) \qquad \nonumber \\
- \frac{1}{L^2}\int _{\rm BZ} d^2k \left[ (t+ \xi V_1)^2 |A_\mathbf{k}|^2 + \left(3V_1 \rho - 2 V_2 |\chi| g_\mathbf{k}(0) \right)^2 \right]^{1/2}
\label{eq:total_free_energy_simplified}
\end{align}
where the momentum-dependent function 
\begin{equation}
g_{\mathbf{k}}( \Theta): = \sin (\mathbf{k}\, \cdot \mathbf{a}_1 + \Theta ) + \sin(\mathbf{k}\, \cdot \mathbf{a}_2 + \Theta )  - \sin (\mathbf{k}\, \cdot (\mathbf{a}_1+\mathbf{a}_2) - \Theta )
\label{eq:g_k}
\end{equation}
is related to $f_\mathbf{k}(\Theta)$ of Eq.~(\ref{eq:f_k}) via $g_{\mathbf{k}}( \Theta) = f_{\mathbf{k}}( \Theta - \pi/2)$.

\subsection{\label{subsec:mean_field_phase_diagram}Mean-field phase diagram of the complete Hamiltonian}

Let us now determine the mean-field phase diagram. Minimization of the free energy (\ref{eq:total_free_energy_simplified}) with respect to $\xi$, $\rho$ and $|\chi|$ yields the coupled set of self-consistency equations
\begin{align}
\label{eq:xi_self_consistent}
\xi = & \,\frac{1}{6L^2}\int_{\rm BZ} d^2k\, \left( |A_\mathbf{k}|^2 (t+ \xi V_1) V_1 \right) / W_{\mathbf{k}} \\
\label{eq:rho_self_consistent}
\rho = & \,\frac{1}{2L^2}\int_{\rm BZ} d^2k\, \left( 3 V_1 \rho - 2 V_2 |\chi| g_{\mathbf{k}}(0) \right) / W_{\mathbf{k}} \\
\label{eq:chi_self_consistent}
|\chi| = & \, - \frac{1}{6L^2}\int_{\rm BZ} d^2k\, \left( 3 V_1 \rho - 2 V_2 |\chi| g_{\mathbf{k}}(0) \right) g_{\mathbf{k}}(0) / W_{\mathbf{k}}
\end{align}
with the dominator
\begin{equation}
W_{\mathbf{k}} = \left[ (t + \xi V_1)^2 |A_\mathbf{k}|^2 + \left( 3 V_1 \rho - 2 V_2 | \chi | g_{\mathbf{k}}(0) \right)^2 \right]^{1/2}
\end{equation}
in the integrand of the three equations. Whereas in principle all three equations have to be solved simultaneously, it is a good approximation to first treat the (relatively weak) effect of the renormalization of the bare hopping amplitude $t$, described by $\xi$, induced by the nearest-neighbor $V_1$-interactions independently, before dealing with the equations for $\rho$ and $|\chi|$, which are order parameters. 

\begin{figure}[t]
	\includegraphics[width=0.65\columnwidth]{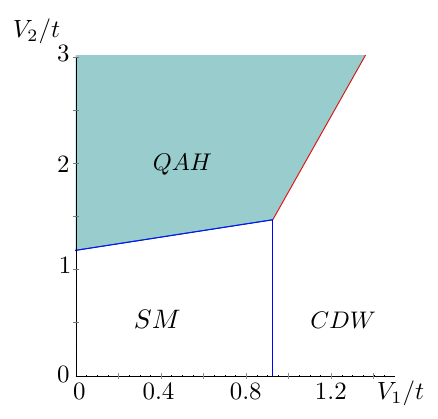}\hspace{0.8cm}
	\caption{Phase diagram as determined from the free energy (\ref{eq:total_free_energy_simplified}) obtained in MFT approximation. The blue lines correspond to second-order phase transitions from a semi-metallic (SM) phase at small interaction values to a CDW phase and a QAH phase. The red line describes a first order phase transition between the CDW and the QAH phase. The green color indicates the region of parameter space, where a non-zero Chern number clearly indicates the non-trivial topological character of the QAH phase, as we will discuss in Sect.~\ref{subsec:Chern_numbers}.}
	\label{fig:phase_diagram}
\end{figure}

\begin{figure}[t]
	\includegraphics[width=\columnwidth]{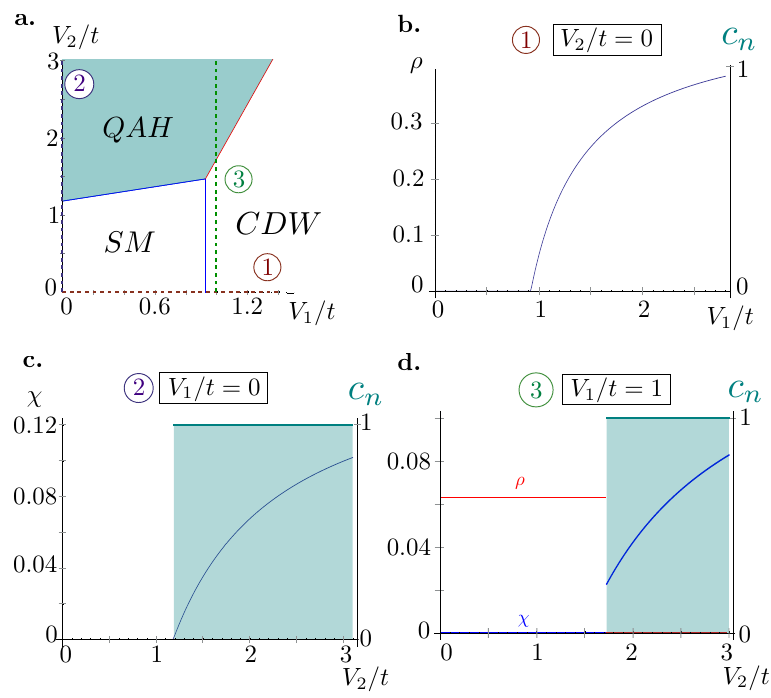}\hspace{0.8cm}
	\caption{Details of the continuous and first-order phase transitions between the SM, CDW and QAH phases found in the phase diagram of Fig.~\ref{fig:phase_diagram}). a) Different cuts in parameter space along the paths 1 and 2 along the $V_1$ and $V_2$ axes (parts a) and c)) and along the path 3 from the CDW into the QAH phase (part d)). Green shading indicates parameter regions where a non-zero Chern number was found.}
	\label{fig:phase_transitions}
\end{figure}

In the region where interactions are weak and the system is in the SM phase ($\rho = 0, \chi=0)$, from Eq.~(\ref{eq:xi_self_consistent}) -- or equivalently (\ref{eq:xi_equation_H1MF}) -- we find for the renormalization of the hopping amplitude, $t' = t + \xi V_1$,
\begin{equation}
\xi = \frac{1}{6L^2}\int_{\rm BZ} d^2k\, |A_\mathbf{k}| = 0.262.
\end{equation}
Next, we consider the limit with only repulsive nearest-neighbor interactions ($V_1$ finite, $V_2 = 0$), for which the interactions favor the formation of a CDW phase. Indeed, from Eq.~(\ref{eq:rho_equation_H1MF}) we obtain
\begin{equation}
\frac{t'}{V_{1c}} = \frac{3}{2L^2} \int_{\rm BZ} d^2k\, |A_\mathbf{k}|^{-1} = 1.345 
\label{eq:V_1c_with_renormalized_hopping}
\end{equation}
and find that at a critical interaction strength of 
\begin{equation}
V_{1c}/t = 0.924
\label{eq:V_1c/t_value}
\end{equation}
in terms of the \textit{bare} hopping amplitude, the system undergoes a second-order phase transition from the SM into a gapped CDW phase. Our results for the mean-field phase diagram, which we will derive and discuss step by step in the following, are summarized in Fig.~\ref{fig:phase_diagram}. Figure \ref{fig:phase_transitions} reveals details of the transitions between the different detected phases of this phase diagram, along three different paths in parameter space which are sketched in Fig.~\ref{fig:phase_transitions}a. For instance, Fig.~\ref{fig:phase_transitions}b, which has been obtained by solving the self-consistency equation (\ref{eq:xi_equation_H1MF}) for $\rho$ for different $V_1$ values, shows the buildup of CDW order $\rho \neq 0$ for $V_1 > V_{1c}$ and makes manifest the second-order nature of the SM to CDW transition. 

Equivalently, we can consider the limit $V_1=0$ in which case we obtain
\begin{equation}
\frac{t'}{V_{2c}} = \frac{1}{3L^2} \int_{\rm BZ} d^2k\, g^2_\mathbf{k}(0) / |A_\mathbf{k}| = 0.845.
\label{eq:V_2c/t}
\end{equation}
For $V_1=0$, where $t'=t$, the critical point 
\begin{equation}
V_{2c}/t = 1.183
\end{equation}
marks a second-order phase transition from the SM to a QAH phase, characterized by a non-zero order parameter $|\chi| \neq 0$. This transition behavior is shown in Fig.~\ref{fig:phase_transitions}c, which has been obtained from the solution of Eq.~(\ref{eq:xi_equation_H1MF}) for different $V_2$ values, and $V_1=0$.

We have also quantitatively studied the build-up of CDW and QAH order in the vicinity of the critical points, yielding -- within numerical accuracy -- a linear behavior according to
\begin{align}
\rho (V_1) = 0.904 \,(V_1 -V_{1c})/t \qquad \text{ for }V_2 = 0 \,\,\,\& \,\,\,V_1 > V_{1c}, \\
|\chi| (V_2) = 0.132\, (V_2- V_{2c})/t \qquad \text{ for }V_1 = 0 \,\,\,\& \,\,\,V_2 > V_{2c}.
\end{align}
when crossing the phase transition lines along the paths 1) and 2) shown in Fig.~\ref{fig:phase_transitions}a.

Let us now comment on the earlier-mentioned controversy on the critical values of the second-order phase-transitions: our values for the renormalization of the hopping $\xi$ and the critical interaction strengths $V_{1c}$ and $V_{2c}$, at which the transitions from the SM to the CDW and QAH phases set in, are in agreement with the values found by Weeks and Franz\cite{Weeks-prb-81-085105}. Our calculations confirm their critical values despite the fact that we work with a spatially reduced mean-field ansatz involving a simpler two-site basis cell as the elementary unit instead of a six-site basis cell\cite{Weeks-prb-81-085105}. Our two-site ansatz is, up to the effect of the renormalization of the hopping amplitude $t' = t + \xi V_1$ which we include in our treatment, equivalent to the ansatz chosen by Raghu \textit{et al.}\cite{Raghu-prl-100-156401}. If we neglect this hopping renormalization our Eq.~(\ref{eq:V_1c_with_renormalized_hopping}) yields $t'/V_{1c} \rightarrow t/V_{1c} =1.345$ from which we find a value $V_{1c}/t=0.743$: this result is a factor of two smaller than the value found by Raghu \textit{et al.}, thus suggesting a problem of double counting of terms in the mean-field treatment as the origin of this discrepancy. We finally would like to remark that it is physically reasonable that the inclusion of the renormalization of the hopping due to the nearest-neighbor interactions $V_1$ produces a larger value for the SM - CDW transition point ($V_{1c}/t = 0.924$ with hopping renormalization vs.~$V_{1c}/t = 0.743$ without this renormalization): the $V_1$-induced increased hopping $t'$ "acts in favor" of the SM phase and thereby pushes the transition point to the CDW phase towards larger $V_1$ values. On the other hand, a moderate value of $V_{1c}/t = 0.924$ compared to the results found by Raghu \textit{et al.} is good news from a  quantum simulation perspective (see Sect.~\ref{sec:qs_Rydberg_atoms} below) as this value suggests that the CDW phase should be accessible already at more moderate values of effective nearest-neighbor interactions. 

Eq.~(\ref{eq:V_2c/t}) not only yields the critical value for $V_1=0$, but also describes the phase boundary between the SM region and the QAH phase for non-zero $V_1$-values,
\begin{equation}
V_{2c}/t = 1.183 + 0.310 \, V_1/t.
\end{equation}
Here, the non-zero slope of $V_{2c}$ as a function of $V_1$ is a result of the renormalized hopping amplitude, which is increased for finite $V_1$ values, thereby stabilizing the SM phase and pushing the QAH phase entrance point to larger $V_2$ values. On the other hand, for finite $V_2 \neq 0$ the phase boundary between the SM region and the CDW ordered phase results to be a vertical line starting off at the critical value $V_{1c}/t = 0.9244$ (see Eq.~(\ref{eq:V_1c/t_value})). This behavior is rooted in the employed ansatz, where the $3(V_1 -2V_2)\rho$ and $3(V_1 -2V_2)\rho^2$ terms in the free energy are approximated by $3 V_1 \rho$ and $3 V_1 \rho^2$, respectively, as justified above in Sect.~\ref{subsec:complete_MF_Ham}. Thus, within this ansatz any (weak) $V_2$-dependence of the $V_1$-driven SM to CDW transition is neglected.

\subsection{\label{subsec:free_energy_and_Chern_numbers} Free energy study}

Obtaining the phase diagram at arbitrary interaction values, and in particular the study of the border between the CDW region and the QAH phase is more intricate as it requires the numerical solution of the coupled set of self-consistency equations (\ref{eq:rho_self_consistent}) and (\ref{eq:chi_self_consistent}). From a technical point of view, numerically searching and finding solutions of the coupled self-consistency equations does generally not preclude the detected extremal points from corresponding to either (meta-stable) saddle-points or \textit{local} minima in the free-energy landscape. It is therefore convenient to work directly with the free energy functional: In contrast to the self-consistency equations based on \textit{local} vanishing of a set of partial derivatives -- the free energy captures the \textit{global} features of the system. It allows to detect and study first order and second order phase transitions in a unified way.

\begin{figure}[t]
	\includegraphics[width=0.95\columnwidth]{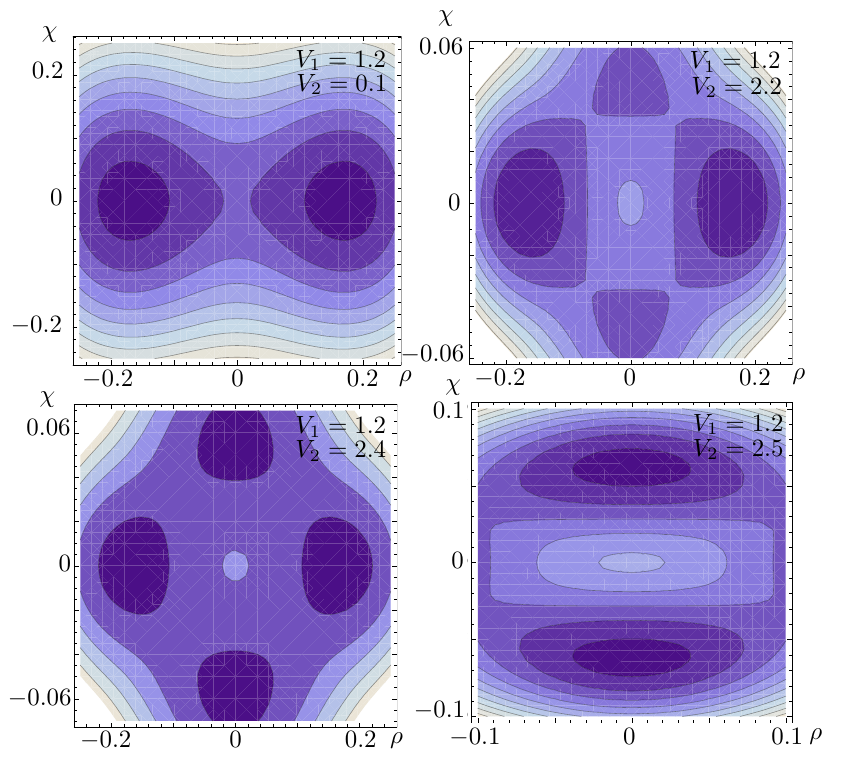}\hspace{0.8cm}
	\caption{Contour plots of the free energy (\ref{eq:total_free_energy_simplified}) for a fixed value of $V_1/t = 1.2$ and increasing values of $V_2$. At small values of $V_2$, the system is in the CDW phase. At increasing values for $V_2$, a second pair of \textit{local} free energy minima at finite $|\chi|$ values appears. At a certain $V_2$ value these minima become global minima and the system undergoes a first-order phase transition from the CDW to the QAH phase.}
	\label{fig:freemethod}
\end{figure}

In particular, the presence of a first-order phase transition between the CDW and the QAH phase as well as its underlying mechanism become especially transparent in a study of the free energy (\ref{eq:total_free_energy_simplified}). Figure \ref{fig:freemethod} shows four contour plots which show the free energy landscape over the $\rho$-$\chi$-plane, for a fixed nearest-neighbor interaction strength of $V_1/t = 1.2$ and increasing values of next-to-nearest neighbor interactions $V_2$. For a small value $V_2/t= 0.1$ the system is in the CDW phase, where the symmetry between $\phi-$ and $\psi$-sublattice sites is spontaneously broken and the system resides in one of the two possible states with either positive or negative relative fermion density imbalance $\pm \rho$ on the two sublattices. These CDW order configurations correspond to the two global minima of the free energy, located on the $\rho$-axis (upper left part of Fig.~\ref{fig:freemethod}). For larger $V_2$ values (upper right part of Fig.~\ref{fig:freemethod}), a second pair of \textit{local} minima of the free energy located on the $\chi$-axis appears. Despite the fact that the system is still in the CDW phase, the presence of these additional local minima at finite $\pm |\chi|$ values is a pre-cursor of the QAH phase and indicates that one approaches the transition into the topologically non-trivial phase. By further increasing $V_2$ one reaches the critical value $V_{2c}$ (lower left part of Fig.~\ref{fig:freemethod}) corresponding to the transition point at which the \textit{local} minima with non-zero $\chi$-value become \textit{global} minima of the free energy and the system enters the QAH phase (lower right part of Fig.~\ref{fig:freemethod}). This clearly indicates that the transition is of first order, which can also be seen from the abrupt jump of the $\rho$- and $\chi$-order parameter values across the transition, as shown in Fig.~\ref{fig:phase_transitions}d.

\begin{figure}[t]
	\includegraphics[width=0.95\columnwidth]{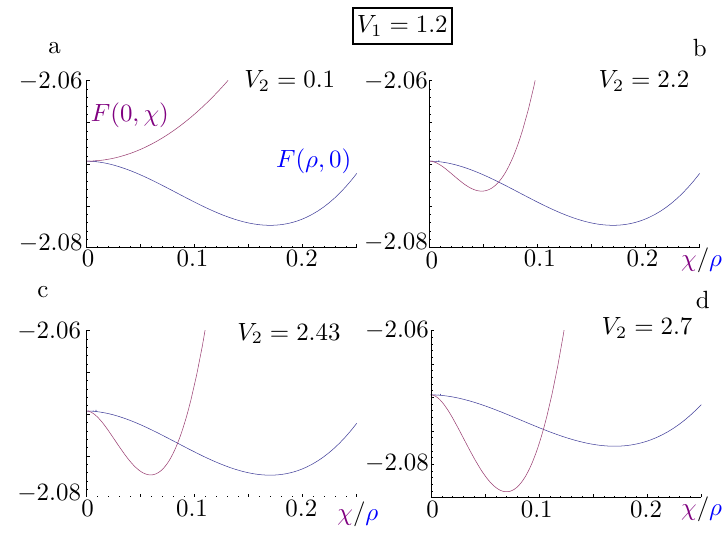}\hspace{0.8cm}
	\caption{Cuts of the free energy (\ref{eq:total_free_energy_simplified}) as a function of $\rho$ and $|\chi|$ for fixed $V_1/t$ and increasing values of $V_2/t$. At small values of $V_2$, the system is in the CDW phase. At larger values of $V_2$, a second \textit{local} free energy minimum appears. At a certain $V_2$ value (around $V_2/t=2.43$) the latter becomes the \textit{global} minimum and the system undergoes a first-order transition from the CDW to the QAH phase.}
	\label{fig:freeecut}
\end{figure}

This transition behavior with the associated change of local into global free energy minima and vice versa is also visible in Fig.~\ref{fig:freeecut}, which shows cuts of the free energy landscape along the $\rho$-axis (blue curve) and the $|\chi|$-axis (red curve). A comparison of the relative depth of the two types of minima allows one to precisely determine the numerical critical values $V_{2c}$, which results in the red line denoting the first-order CDW to QAH transition shown in Fig.~\ref{fig:phase_diagram}. One can show that for a fixed $V_1$ interaction strength there exists only a single point $V_{2c}(V_1)$ at which the depth of the two types of free energy minima is the same, which in other words implies that the CDW-QAH transition is \textit{sharp} and that a co-existence of the topologically trivial CDW phase with the QAH phase in a finite region of parameter space is precluded, at least when working with the mean-field free energy expression (\ref{eq:total_free_energy_simplified}). 

\subsection{\label{subsec:Chern_numbers} Chern number characterization}

To unequivocally establish the topologically non-trivial character of the QAH phase, we have calculated the Chern numbers characterizing the topological nature of the system in the various regions of the phase diagram, which for the Hamiltonian of interest in this work has -- to the best of our knowledge -- not been done so far. In contrast to the $\chi$-order parameter, which is a local order parameter, the Chern numbers are topological invariants and can be regarded as \textit{non-local} order parameters, which are only sensitive to the global properties of the system. They characterize the electronic states in the TMI phase of QAH type. As the ground state of the interacting QAH phase breaks time-reversal symmetry, this is described by a $U(1)$ Chern number\cite{qi-rmp-83-1057}.

Within the MFT approximation the Hamiltonian (\ref{eq:complete_mean_field_Ham_momentum_space}) is a quadratic form and can be readily diagonalized in momentum space, and its momentum-dependent energy eigenvalues and eigenstates can be given analytically. We emphasize that despite this fact it is non-trivial to calculate the corresponding Chern numbers. The reason is that a numerical integration over the momenta of the Brillouin zone generically in one way or the other requires a discretization and the evaluation of the eigenfunctions and derivatives at a particular discrete set of momenta. It is a well-known problem by now\cite{bib:fukui} that if this discretization is done naively, the gauge invariance of the local curvature over the Brillouin zone is lost and the corresponding numerically obtained approximate Chern numbers fail to be an integer anymore, thereby loosing their intrinsic topological character. Thus, to correctly obtain the Chern numbers which distinguish the topologically trivial and non-trivial phases of the system, we employ an efficient numerical algorithm which has been proposed by Fukui \textit{et al.}\cite{bib:fukui}. The distinguishing feature of the method is that by construction of the algorithm the obtained Chern numbers are manifestly gauge-invariant and integer-valued, even when one works with a discretized Brillouin zone. The method only requires as an input the wave function of the system at such a finite set of points, without the need of working in a particular gauge or specifying gauge-fixing conditions. In addition, it produces the correct Chern numbers even for a fairly coarse-grained discretization of the Brillouin zone \cite{bib:fukui}. We refer the reader to Appendix \ref{app:Chern_numbers} for more details on the numerical calculation of the Chern numbers in this work. 

Our numerical analysis confirms the topological QAH character of the $\chi \neq 0$ phase, as it unequivocally yields a Chern number value equal to one for the $\chi$-phase region and, as expected, vanishing Chern numbers for the topologically trivial SM and CDW ordered phases (see Figs.~\ref{fig:phase_diagram} and \ref{fig:phase_transitions}c \& d). We emphasize that while the order parameter $\chi$ is local in the QAH phase, the Chern number is a non-local observable that provides additional information about the nature of this exotic phase.

\subsection{\label{subsec:coexistence} Phase coexistence}

Our derivation of the phase diagram and the characterization of the topologically trivial and non-trivial phases in the previous sections is based on the spatially homogeneous mean-field decoupling with a two-site basis cell as the elementary unit of the hexagonal lattice and two variational parameters $\rho$ and $|\chi|$, quantifying CDW order and the QAH phase. Within the employed reduced ansatz - previously also used by Raghu \textit{et al.}\cite{Raghu-prl-100-156401}, the absolute value of the complex QAH order parameter has been set to an equal value on $\phi$- and $\psi$-sublattice sites, and the phases have been fixed according to Eq.~(\ref{eq:chi_parameters_reduced}). 

A spatially extended, more sophisticated mean-field ansatz can give rise to a richer phase diagram: In previous work Weeks and Franz\cite{Weeks-prb-81-085105} showed that within a mean-field ansatz based on an elementary six-site basis cell the system assumes, in addition to the SM, the CDW and the QAH phases, a so-called Kekul\'e phase over an extended region in the parameter space.

Instead of resolving more phases, here we are interested in the complementary question whether a more general \textit{two-site} mean-field ansatz, which allows for arbitrary values of $\chi_\phi$ and $\chi_\psi$, enables the coexistence of a topologically trivial (CDW) with a non-trivial (QAH) phase within a finite region of the $V_1$-$V_2$-parameter space. In the preceding sections, we have shown that there is no phase coexistence within the reduced ansatz specified by Eq.~(\ref{eq:chi_parameters_reduced}). 
To study the question of phase coexistence within a two-site ansatz, which allows for arbitrary values of $\chi_\phi$ and $\chi_\psi$, we start with the free energy 
\begin{align}
\label{eq:total_free_energy_general_ansatz}
F/N_\Lambda =  & \,\,3V_1 \xi^2 + 3 V_1 \rho^2  + 3 V_2 \left( | \chi_\phi |^2 +  | \chi_\psi |^2 \right) \\
& - V_2 \frac{1}{L^2}\int _{\rm BZ} d^2k \left( | \chi_\phi | f_{\mathbf{k}}(\Theta_\phi)+  | \chi_\psi | f_{\mathbf{k}}(\Theta_\psi) \right) \nonumber \\
& -  \frac{1}{L^2}\int _{\rm BZ} d^2k \left[ (t+ \xi V_1)^2 |A_\mathbf{k}|^2 \right. \nonumber \\
& \,\,\,\left. + \left( 3 V_1 \rho + V_2  \left( | \chi_\phi | f_{\mathbf{k}}(\Theta_\phi) - | \chi_\psi | f_{\mathbf{k}}(\Theta_\psi) \right) \right)^2 \right]^{1/2}\text{,} \nonumber
\end{align}
which has been obtained from Eq.~(\ref{eq:total_free_energy}) by neglecting the constant density term and the $V_2$-dependence in the terms proportional to $\rho$ and $\rho^2$ (see discussion in Sect.~\ref{subsec:complete_MF_Ham}). First, we note that due to the identities
\begin{align}
& f_\mathbf{k}(\Theta) =\cos(\Theta) f_\mathbf{k}(0)-\sin(\Theta) g_\mathbf{k}(0), \\
& \int _{\rm BZ} d^2k f_\mathbf{k}(\Theta) = \int _{\rm BZ} d^2k g_\mathbf{k}(\Theta) = 0,
\end{align}
for the momentum-dependent functions $f_\mathbf{k}$ and $g_\mathbf{k}$ defined in Eqs.~(\ref{eq:f_k}) and (\ref{eq:g_k}) the first integral in the free energy (\ref{eq:total_free_energy_general_ansatz}) vanishes identically\footnote{Eq.~(50) follows from the symmetry properties of the Brillouin zone, as discussed in App.~\ref{app:brillouin_zone}, and can be readily shown by analytically carrying out the integration e.g.~over the rectangular-shaped Brillouin zone, as depicted in the right part of Fig.~\ref{fig:bzv3}.}. As above, we determine the ground state diagram by numerically minimizing the free energy Eq.~(\ref{eq:total_free_energy_general_ansatz}) -- see also Sect.~\ref{subsec:free_energy_and_Chern_numbers}. Due to the enlarged number of variational parameters (absolute values and phases of $\chi_\phi$ and $\chi_\psi$, as well as $\rho$) in the more general ansatz, we use a simulated annealing routine to determine the absolute minimum of the free energy.

For $V_1=0$, inspection of the self-consistency equations derived from Eq.~(\ref{eq:total_free_energy_general_ansatz}) shows that $|\chi_\phi | - |\chi_\psi |= 0$ if and only if $\Theta_\phi - \Theta_\psi = \pi$. As expected, in this case the numerical minimization reproduces the transition from the SM region to the QAH phase with a non-zero $\chi$-order parameter according to Eq.~(\ref{eq:chi_parameters_reduced}). On the other hand, for the SM to CDW transition, we find that the $\chi$ - expectation values are real and opposite in sign, however, the absolute minimum of the free energy corresponds to the CDW-ordered phase with a non-zero $\rho$-expectation value and a real-valued (renormalized) nearest-neighbor and a real-valued next-to-nearest-neighbor hopping amplitude. 

To study the question of phase coexistence, we have analyzed in detail the transition from the CDW phase (at fixed $V_1$ values) into the QAH phase. Potential coexistence of a CDW-ordered and a QAH phase in a finite interval $[ V_{2,\mathrm{min}}, V_{2,\mathrm{max}} ]$ (depending on the chosen $V_1$ value) could correspond to (i) the coexistence of two global free-energy minima, namely one with $\rho \neq 0$ and another with $\chi_\phi \neq 0, \, \chi_\psi \neq 0$, or (ii) the appearance of a global minimum of the free energy with both $\rho \neq 0$ and $\chi_\phi \neq 0, \, \chi_\psi \neq 0$. A careful numerical analysis along paths of increasing $V_2$-values, as described in Sect.~\ref{subsec:free_energy_and_Chern_numbers} and illustrated in Fig.~\ref{fig:freeecut}, excludes both possibilities.

We conclude that even within a generalized two-site MFT ansatz the coexistence of a topologically trivial and a non-trivial phase is precluded in this model. On the one hand, these findings thereby provide an a posteriori justification for using the reduced MFT ansatz with a single parameter $|\chi|$ according to the choice of Eq.~(\ref{eq:chi_parameters_reduced}) in our MFT analysis and previous studies\cite{Raghu-prl-100-156401}. On the other hand, in view of a potential experimental observation of the predicted phases, to be discussed in detail in the next section, these results suggest a \textit{sharp} phase boundary between the CDW-ordered and the QAH phase.

\section{\label{sec:qs_Rydberg_atoms} Quantum simulation with Rydberg atoms in optical lattices}

The dynamics of cold atoms in optical lattices is governed by Hubbard Hamiltonians \cite{PhysRevLett.81.3108,greiner-nature-415-39,jaksch-ann-phys-315-52,bloch-rmp-2008}. The ability to control the underlying interactions and Hamiltonian parameters as well as the dimensionality of the setup and the type of the lattice, in combination with sophisticated state preparation and measurement techniques, have turned cold atom experiments into a versatile platform for the study a wide range of complex many-body quantum systems, described by models originating, e.g.,~from the field of condensed-matter \cite{lewenstein-adv_amo_phys-56-135,lewenstein-book} or motivated from high-energy physics \cite{arxiv_Banerjee_2012}. Specifically, as we will outline in detail in this section, cold fermionic atoms, loaded in an optical lattice and weakly laser-coupled to Rydberg states, provide a platform for a reliable quantum simulation of the Hamiltonian of interest (see Eq.~(\ref{hamiltonian})). 

Whereas the hopping dynamics in a two-dimensional honeycomb lattice geometry can be realized and controlled by a combination of sinusoidal optical trapping potentials\cite{duan-prl-91-090402,zhu-prl-98-260402,arxiv_alba-1107.3673-2011}, the main challenge lies in the implementation of the repulsive nearest- and next-to-nearest neighbor interactions of strength $V_1$ and $V_2$ between fermions residing at different lattice sites, at distances $d_{\mathrm{nn}} = a$ and $d_{\mathrm{nnn}} = \sqrt{3}a$, respectively. Here we propose to use laser dressing of ground state atoms with strong and long-range Rydberg interactions to induce effective interatomic interactions over distances of several lattice sites. In particular we are interested in the parameter regime $V_1, V_2 > t$ and $V_2 \sim V_1$, which is essential to be able to explore the predicted quantum phases beyond the SM region of the phase diagram. 

Atoms laser-excited to high-lying Rydberg states\cite{gallagher-book,saffman-rmp-82-2313} interact strongly via long-range dipole-dipole or van-der-Waals interactions over distances of several $\mu$m. The interactions as well as the electronic lifetimes of atoms in Rydberg states are several orders of magnitude larger than for atoms in electronic ground states. One of the hallmark features of the strong Rydberg interactions is the "dipole blockade" effect, where large electronic level shifts inhibit the laser-excitation of more than a single atom to a Rydberg state within a certain blockade radius. In a many-body context, this mechanism gives rise to collective effects and strongly correlated coherent many-body dynamics in laser-driven atomic gases \cite{raitzsch-prl-100-013002,tong-prl-93063001,singer-prl-93-163001,cubel-pra-72-023405,vogt-prl-97-083003,mohapatra-prl-98-113003,heidemann-prl99-163601,sun-njp-10-045032,weimer-prl-101-250601,pohl-prl-104-043002,olmos-prl-103-185302}. As suggested more than a decade ago \cite{jaksch-prl-85-2208,lukin-prl-87-037901} and recently demonstrated in the laboratory \cite{wilk-prl-104-010502,isenhower-prl-104-010503} internal state-dependent switchable Rydberg interactions can be exploited to realize neutral-atom entangling quantum gates, which are a key ingredient for the development of scalable cold atom quantum information processing architectures. In the field of quantum simulation the availability of (many-particle) Rydberg gates\cite{moller-prl-100-170504,mueller-prl-102-170502} allows one to realize a digital quantum simulation architecture, where coherent (as well as dissipative) time evolution of open many-body quantum systems is realized by stroboscopic sequences of gate operations \cite{weimer-nphys-6-382,weimer-quantuminfprocess-10-885}. On the other hand, in the direction of analog quantum simulation, the ability to tune the strength and form of the interactions by means of static and time-dependent fields allows one to engineer a broad range of Hamiltonians with effective spin-spin interactions \cite{pupillo-prl-104-223002,schachenmayer-njp-12-103044,wuester-njp-13-073044,lesanovsky-prl-106-025301,henkel-prl-104-195302}.

\subsection{\label{subsec:laser_dressing_scheme} Rydberg dressing scheme and effective long-range interactions}
Following the analog quantum simulation approach, in the envisioned implementation, we aim at inducing effective interatomic interactions by the application of "always-on" and global, i.e.~spatially homogeneous CW-laser fields, without the requirement of single-site addressability, as recently achieved by several groups \cite{bakr-science-329-547,weitenberg-nature-471-319}. To convey the main idea, we start by outlining first the ideal simulation scheme and comment afterwards on the range of validity, imperfections and the experimental feasibility of our approach (see Sects.~\ref{subsec:validity} and \ref{subsec:parameters} below). 

\begin{figure}[t]
	\includegraphics[width=0.9\columnwidth]{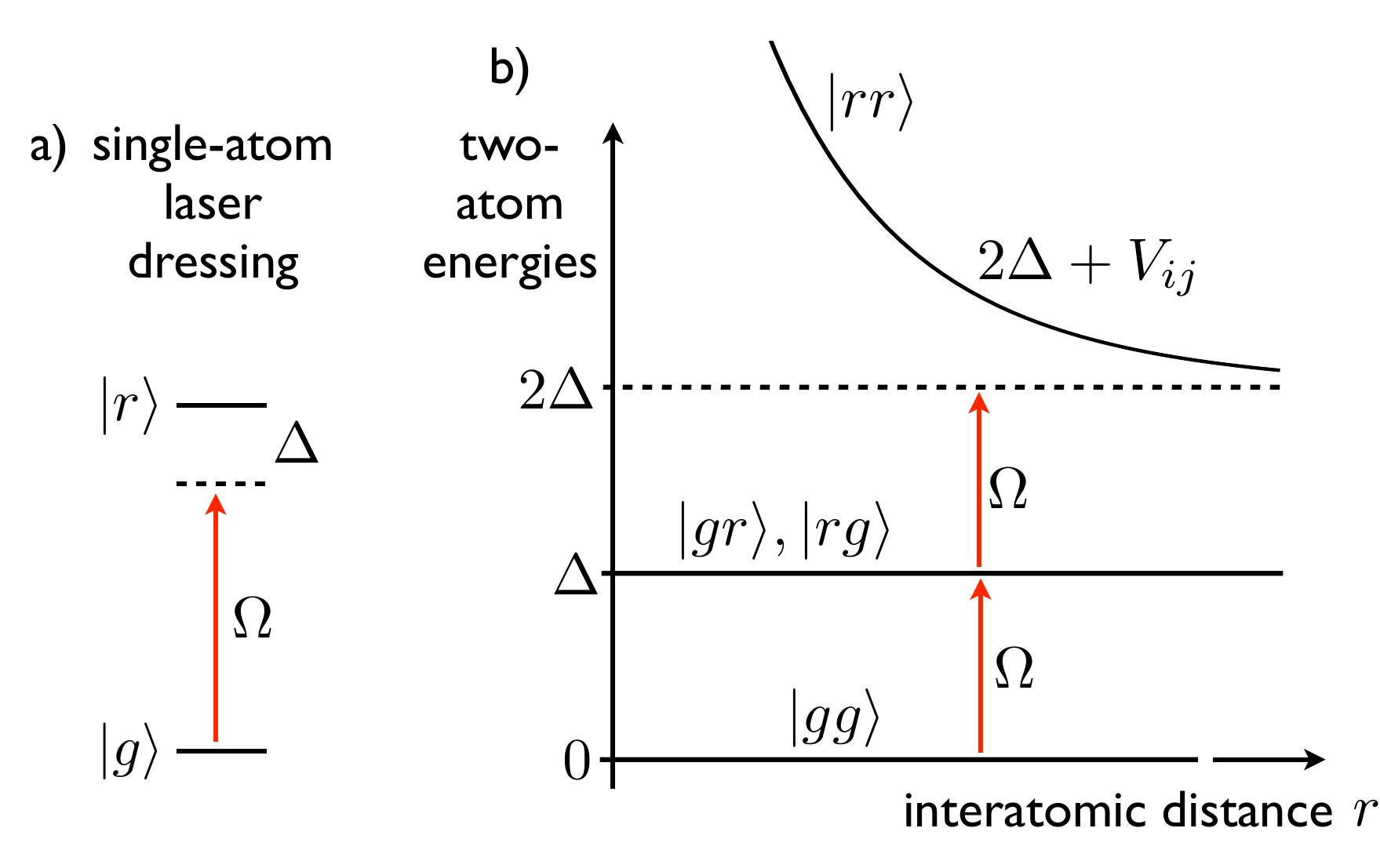}\hspace{0.8cm}
	\caption{a) Atoms residing in the stable electronic ground state $\ket{g}$ are coupled far off-resonantly ($\Delta \gg \Omega$) and in a red-detuned way ($\Delta > 0$) to the meta-stable Rydberg state $\ket{r}$. b) Schematics of the two-particle energy surfaces corresponding to the four two-atom electronic states; laser couplings between the electronic states are indicated by red arrows. The state $\ket{rr}$ with both atoms excited to the Rydberg state has a distance-dependent additional two-body energy shift due to repulsive van-der-Waals interactions $V_{ij}$.}
	\label{fig:RydbergFigure1}
\end{figure}

We consider fermionic atoms loaded in a hexagonal optical lattice, where - in addition to the external dynamics described by the hopping part of Hamiltonian (\ref{hamiltonian}) - the internal dynamics of the atoms is governed by a Hamiltonian 
\begin{equation}
H = \sum_i H_i + \sum_{i<j} H_{ij}.
\end{equation}
Here, in the co-rotating frame (and in rotating-wave approximation) the single-atom laser-driving term
\begin{equation}
\label{eq:H_i_laser_Ham}
H_i = (\Omega \ket{r}\bra{g}_i + h.c.) + \Delta \ket{r} \bra{r}_i
\end{equation}
accounts for coupling of atoms residing in a (meta)stable electronic ground state $\ket{g}$ to an excited Rydberg state $\ket{r}$ with a (two-photon) Rabi frequency $\Omega$ and detuning $\Delta$, as depicted in Fig.~\ref{fig:RydbergFigure1}a. Pairwise interactions between atoms excited to the Rydberg state $\ket{r}$ are described by 
\begin{equation}
H_{ij} = V_{ij} \ket{r}\bra{r}_i \otimes  \ket{r}\bra{r}_j.
\end{equation}
We will focus on Rydberg $s$-states with repulsive van-der-Waals interactions $V_{ij} > 0$, in which case the interactions $V_{ij} = C_\alpha |\mathbf{r}_i - \mathbf{r}_j|^{-\alpha}$ (with $\alpha=6$) are isotropic and fall off quickly with the interparticle distance. The pairwise interaction potential curves as a function of the distance between two atoms and the laser coupling between these curves are shown schematically in Fig.~\ref{fig:RydbergFigure1}b. 

\begin{figure}[t]
	\includegraphics[width=0.9\columnwidth]{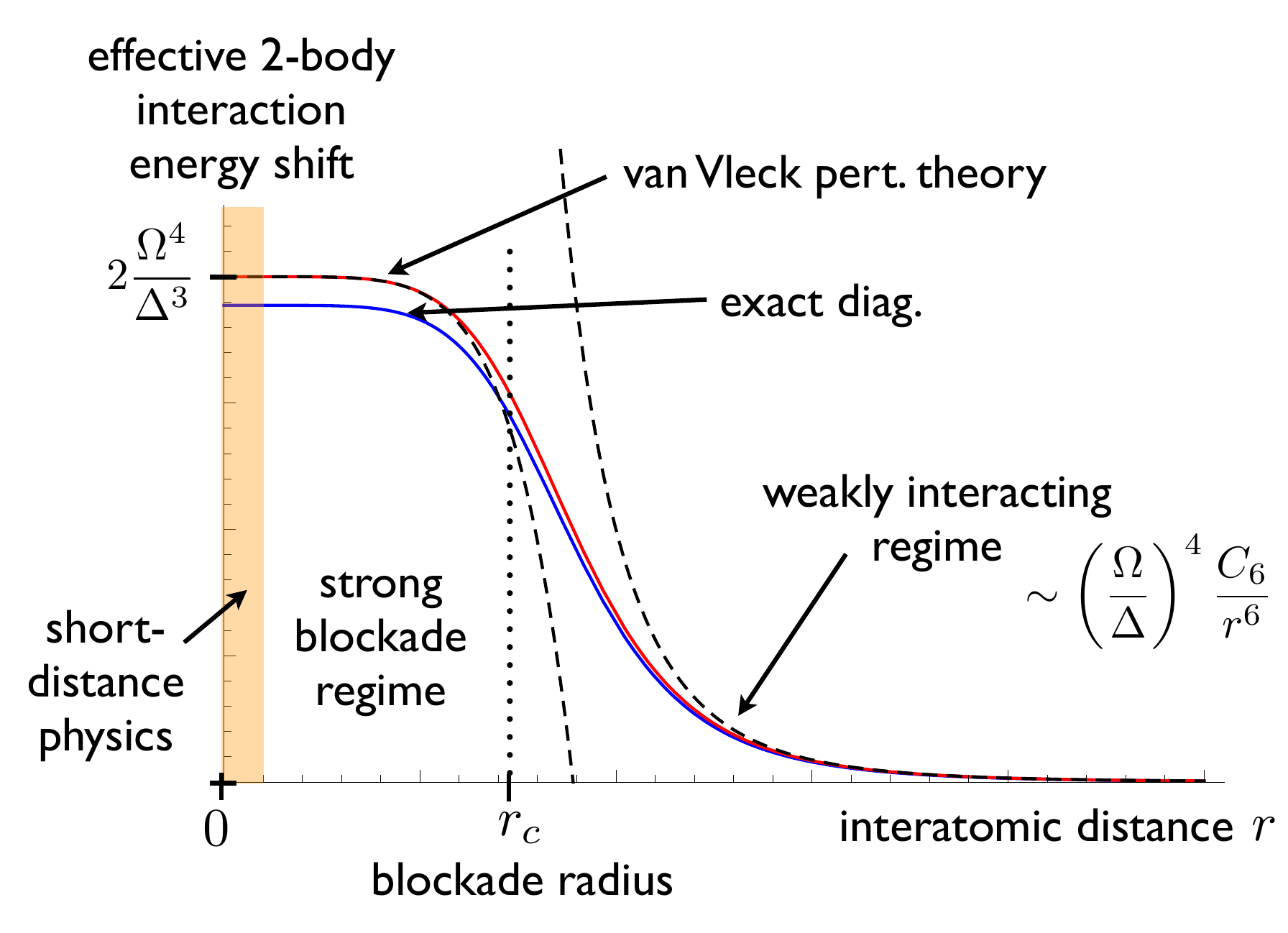}\hspace{0.8cm}
	\caption{Effective two-body interaction potential between Rydberg-dressed ground state atoms. The red curve shows the energy obtained from fourth-order theory (see Eq.~(\ref{eq:real_part})), whereas the blue curve is obtained from exact diagonalization of the two-particle Hamiltonian. At large distances, $r > r_c = (C_6/(2 \Delta))^{1/6}$ (dotted, vertical line as a guide to the eye), Rydberg-dressed ground state atoms interact weakly according to $1/r^6$ van-der-Waals interactions, which are reduced by a factor $(\Omega/\Delta)^4$. In contrast, in the strongly interacting regime, at interatomic distances smaller than $r_c$, the two-body interaction potential exhibits a plateau-like structure, with a distance-independent energy shift $2\, \Omega^4/\Delta^3$. The black dashed lines correspond to the approximate analytical expressions for the two-atom potential in the weakly interacting and the strong blockade limit, as given by Eqs.~(\ref{eq:energy_shift_weakly_interacting}) and (\ref{eq:energy_shift_strong_blockade}), respectively. At even shorter distances \cite{gallagher-book} electronic wave functions of Rydberg atoms start to overlap and interactions are no longer described by van-der-Waals interactions. Parameters for this plots where chosen $C_6 = \Delta = 10 \Omega$.}
	\label{fig:RydbergFigure2}
\end{figure}

For far off-resonant laser coupling, $\Omega/\Delta \ll 1$, and the atoms initially residing in the electronic ground state $\ket{g}$, transitions from the ground state to states with one or more atoms excited to a Rydberg state are energetically not accessible and strongly suppressed. Nonetheless, the laser coupling induces a weak admixture of Rydberg states to the ground state atoms (dressing), which thereby "inherit" some part of the Rydberg interaction character. To quantitatively describe the resulting effective interactions we derive the effective Born-Oppenheimer two-particle potential surface for two Rydberg-dressed ground state atoms by means of standard (Van-Vleck-type) fourth-order perturbation theory\cite{Shavitt-jchemphys-73-5711}, whose validity is based on the small parameter $\Omega/\Delta \ll 1$. The resulting distance-dependent energy shift for two Rydberg-dressed ground state atoms is given by
\begin{equation}
\label{eq:real_part}
(\Delta E)_{\ket{gg}} = 2 \frac{\Omega^4}{\Delta^3} \left[ 1+ \frac{2\Delta}{V_{ij}} \right]^{-1},
\end{equation}
where we have subtracted the trivial, interaction-independent single-particle (fourth order in $\Omega/\Delta$) AC-Stark-shift $2(\Omega^2/\Delta) (1 - (\Omega/\Delta)^2))$ of the two atoms in $\ket{g}$. The effective two-atom interaction potential curve is shown in Fig.~\ref{fig:RydbergFigure2}. 

In the \textit{weakly interacting limit} ($V_{ij} \ll \Delta$), at large interatomic distances, the energy shift (\ref{eq:real_part}) reduces to 
\begin{equation}
\label{eq:energy_shift_weakly_interacting}
(\Delta E)_{\ket{gg}}  = 
\left( \frac{\Omega}{\Delta} \right)^4 V_{ij} \left( 1 - \frac{V_{ij}}{2 \Delta} \right).
\end{equation}
In this regime, both atoms can be partially excited to the Rydberg state, independently of the state of the other atom. As a result, the interaction shift is in leading order given by the bare Rydberg interaction energy $V_{ij} = C_\alpha |\mathbf{r}_i - \mathbf{r}_j|^{-\alpha}$, reduced by the small prefactor $(\Omega/\Delta)^4$, which corresponds to the probability to find both dressed ground state atoms simultaneously in the Rydberg state. On the other hand, in the \textit{strongly interacting limit} ($V_{ij} \gg \Delta$) at small interparticle distances, where no more than one atom can be simultaneously excited to the Rydberg, the energy shift becomes 
\begin{equation}
\label{eq:energy_shift_strong_blockade}
(\Delta E)_{\ket{gg}}  = 2 \frac{\Omega^4}{\Delta^3} \left(1 - 2\frac{\Delta}{V_{ij}}\right).
\end{equation}
Thus, up to a sub-leading correction, the energy shift is given by the distance-independent value of $2 \, \Omega^4/\Delta^3$. The critical length scale $r_c$ at which the crossover from the plateau-like behavior with a universal value for the interaction shift at short distances to the weakly interacting regime takes place, is determined by the condition $2 \Delta \approx V_{ij}$, yielding 
\begin{equation}
\label{eq:blockade_radius}
r_c = (C_\alpha /2 \Delta)^{1/\alpha}.
\end{equation}

The key idea to access the different regions of the phase diagram shown in Fig.~\ref{fig:phase_diagram} in the cold atom quantum simulation lies in controlling the relationship of the nearest- and next-to-nearest neighbor distances $d_{\mathrm{nn}}$ and $d_{\mathrm{nnn}}$, which are naturally fixed by the lattice geometry, and the critical length scale $r_c$, which in turn is determined by the choice of the electronic Rydberg state $\ket{r}$ and the tunable laser parameters: To access the CDW phase it is sufficient to assure $r_c < d_{\mathrm{nnn}}$, which implies $V_1 > V_2$. Exploring the region where the existence of the QAH phase is expected, as well as its vicinity in parameter space, is more challenging: the crucial point is to realize a situation where pairs of nearest and next-to-nearest atoms interact strongly, i.e. are located within the blockade radius ($d_{\mathrm{nn}} < r_c$ and $d_{\mathrm{nnn}} < r_c$), where due to the plateau-like behavior of the potential curve (see Fig.~\ref{fig:RydbergFigure2}) the interaction strengths are of comparable magnitude, $V_1 \sim V_2$. Due to the continuous decrease of the Rydberg interactions as a function of the interparticle distance inevitably some small, effective longer-range interaction terms (strength $V_3$ for atoms separated by a distance $2a$, $V_4$ at a distance $\sqrt{7}a$, etc.) will be induced. However, if these distances of at least $2a$ are larger than the critical radius $r_c$ the strength of residual longer-range interactions $V_3, V_4, \ldots$ fall into the weakly interacting regime, where they decrease rapidly with increasing distance ($\sim r^{-6}$) and are substantially smaller than the desired nearest- and next-to-nearest-neighbor interactions $V_1$ and $V_2$. The question of the qualitative and quantitative influence of these additional long-range interaction terms on the phase diagram is interesting, and it has been partially addressed in recent work \cite{Weeks-prb-81-085105}, where for fermions on a square lattice the authors found that a third-nearest-neighbor repulsive interaction stabilized a QAH phase. However, a more detailed analysis of this question requires either a more sophisticated (i.e.~spatially extended) mean-field ansatz or exact diagonalization calculations, both of which lie beyond the scope of the present work. We note that inclusion of these terms might play a role in determining the precise shape of the phase boundary, at which the system undergoes a transition into the non-trivial topological QAH phase. Whereas the mean-field analysis clearly indicates that strong next-to-nearest interactions are essential for the emergence of the QAH phase, the question whether $V_1 \simeq V_2$ is sufficient to enter the non-trivial topological phase is a question to be ultimately answered by an experiment.

\subsection{\label{subsec:validity} Validity of the effective two-body interaction potential}
In this subsection, we discuss the regime of validity for the effective two-body potential derived in Sect.~\ref{subsec:laser_dressing_scheme}. First of all, the off-resonant laser-coupling to the meta-stable Rydberg state $\ket{r}$ with a finite decay rate $\gamma_r$ induces an effective decay rate $\gamma_\mathrm{eff}$ for the dressed ground state atoms, which limits the description of the interactions in terms of the effective two-body potential (\ref{eq:real_part}) to time scales $\tau \ll \gamma_\mathrm{eff}^{-1}$ on which incoherent decay processes from the Rydberg state are negligible. In the limit $\gamma_r \ll \Omega$, which is well-fulfilled for typical Rydberg states and moderately large Rabi frequencies $\Omega$, for the purpose of a reliable estimate one can simply include an imaginary part $\Delta \ket{r}\bra{r} \rightarrow (\Delta -i \gamma_r/2) \ket{r}\bra{r}$ in the single-particle Hamiltonian $H_i$ (see Eq.~(\ref{eq:H_i_laser_Ham})). For this non-hermitian Hamiltonian which includes the effect of decay from the Rydberg state, the van Vleck-perturbation theory treatment then yields an effective decay rate per atom $\gamma_{\mathrm{eff}}= (\Omega/\Delta)^2 \gamma_r$ in leading order in $(\Omega/\Delta)$, both in the strongly and in the weakly interacting regime. At longer times such undesired decay processes from the Rydberg state set in, events which are typically associated to strong mechanical forces leading to atomic losses and heating of the dressed Rydberg gas, as recently studied in detail by Gl\"atzle \textit{et al.}\cite{arxiv_glaetzle-1207.2659-2012}

In addition, it is a priori not clear that an effective description in terms of effective pairwise two-body interactions is possible, and that three- and higher-order many-body interaction terms are negligible, in particular in the \textit{strongly interacting regime}, where collective effects are expected to play an important role. As pointed out in previous work \cite{honer-prl-105-160404}, the range of validity of the effective two-body description is determined by the interplay of the following conditions: on the one hand, due to the blockade radius $r_c$ the maximal density of Rydberg atoms that can be laser-excited in a two-dimensional lattice setup is given by $n_\mathrm{Ryd}^{\mathrm{max}} = 1 / r_c^2$. On the other hand, at large distances, the condition of weak dressing predicts a Rydberg atom density $n_\mathrm{Ryd} = (\Omega/\Delta)^2 n$, where $n$ is the density of ground state atoms in the lattice. Thus, we find the condition $n r_c^2 \ll (\Delta/\Omega)^2$ that has to be fulfilled. Using the expression for the blockade radius (\ref{eq:blockade_radius}) this amounts to the requirement
\begin{equation}
\label{eq:density_requirement}
n \ll \left( \frac{\Delta}{\Omega} \right)^2 r_c^{-2} = 2^{\frac{2}{\alpha}} \frac{\Delta^{2+\frac{2}{\alpha}}}{\Omega^2 C_\alpha^{\frac{2}{\alpha}}} \,\,\stackrel{\mathrm{vdW}}{=}\, \, 2^{1/3} \frac{\Delta^{7/3}}{\Omega^2 C_6^{1/3}},
\end{equation}
which can in principle always be fulfilled by choosing a sufficiently large detuning $\Delta$ at a given density $n$ of ground state atoms. The ground state atom density is determined by the lattice geometry and the filling factor ($n=1/2$ in the present case). Sixth-order in $(\Omega/\Delta)$ perturbation theory then shows that in the limit specified by (\ref{eq:density_requirement}) -- and even in the strongly interacting regime, $V_{ij}\gg \Delta$ -- interactions are well-described by effective pairwise two-body terms according to Eq.~(\ref{eq:real_part}) and that the strength of the leading-order corrections due to three-body interactions is of the order $\Omega^6/\Delta^5$, i.e., at least two orders of magnitude smaller (in the parameter $\Omega/\Delta \ll1$) than the two-body terms.

\subsection{\label{subsec:parameters} Experimental feasibility, Rydberg states and laser parameters for a quantum simulation}
Very recently, the simulation of (non-interacting) Dirac fermions with a quantum degenerate Fermi gas of $^{40}$K atoms loaded into a tunable honeycomb lattice has been demonstrated experimentally\cite{tarruell-nature-483-302}. On the other hand, several groups have recently achieved trapping and laser excitation of cold atoms in optical lattices into Rydberg states, and reported the observation of the Rydberg blockade in this setup\cite{viteau-prl-107-060402,anderson-prl-107-263001}. If combined in a single experiment, these achievements directly enable the implementation of (i) the hopping dynamics of fermionic atoms in a two-dimensional hexagonal optical lattice potential, along with (ii) the engineering of the effective interactions over several lattice sites by means of the described Rydberg laser-dressing scheme, which are required for the implementation of the complete interacting model Hamiltonian (\ref{hamiltonian}).

An optimal choice of Rydberg states and laser parameters clearly depends on the specific experimental implementation under consideration, the fermionic species used, and technical limitations such as available laser power and achievable two-photon Rabi frequencies for the coherent ground-to-Rydberg state laser excitation. However, let us provide a rough, though conservative estimate for a set of parameters, under which the requirements discussed in Sect.~\ref{subsec:laser_dressing_scheme} are fulfilled and fundamental, non-technical limitations listed in Sect.~\ref{subsec:validity} are small. We assume a real-space lattice spacing $a$ of about 500 nm\cite{tarruell-nature-483-302}. As discussed above, the exploration of the different phases of the model requires the critical radius $r_c$ (\ref{eq:blockade_radius}) to be on the order of the lattice spacing $a$ between neighboring sites. For bare van-der-Waals interactions of about $V \sim 2 \pi \times 100$ MHz for Rydberg $s$-states (angular momentum $l=0$) with a principal quantum number around $n=30$ at this distance\cite{singer-jphysb-38-S295}, this implies a detuning of $\Delta = V/2 = 2 \pi \times 50$ MHz. Assuming a two-photon Rabi frequency of $\Omega =2 \pi \times 5$ MHz yields the smallness parameter $\Omega/\Delta = 0.1$ such the assumptions underlying the perturbative treatment are well-fulfilled. The resulting energy scale of the effective two-body interaction energy shift of dressed ground state atoms is $(\Delta E)_{\ket{gg}} \sim 2 \Omega^4/\Delta^3 = 2\pi \times 10 \,\text{kHz}$. This energy corresponds to the $V_1$ and $V_2$ nearest- and next-to-nearest neighbor interaction strengths of Hamiltonian (\ref{hamiltonian}), thus being comparable to typical tunnel rates $t$ on the order of (a few) kHz\cite{greiner-nature-415-39}.

Regarding imperfections: Rydberg $s$-states with a principal quantum number around $n = 30$ have typical radiative life-times $\tau_r$ of a few tens of microseconds at room-temperature\cite{gallagher-book,saffman-rmp-82-2313}. For $\tau_r \sim 30 \,\mu$s we find an effective decay rate of $\gamma_{\rm eff} = (\Omega/\Delta)^2 \gamma_r = 2 \pi \times 50$\,Hz, which is much smaller than the frequency scale corresponding to the effective interactions $(\Delta E)_{\ket{gg}}$. Finally, at half-filling of the lattice, for the present set of parameters the condition (\ref{eq:density_requirement}) is also fulfilled, such that interactions between laser-dressed ground state atoms are well-described by the effective two-body potentials, and three- and higher-order many-body effects are negligible.

\section{\label{sec:conclusions_and_outlook} Conclusions and Outlook}
In this work, we have studied a system of interacting spinless fermions with nearest- and next-to-nearest neighbor repulsion on the honeycomb lattice and proposed a quantum simulation scheme of the model, based on cold fermionic Rydberg atoms in an optical lattice. In the first part of the manuscript, we have determined the phase diagram of the model in a MFT treatment, yielding topologically trivial SM and CDW ordered phases, as well as the existence of a QAH phase for sufficiently strong next-to-nearest neighbor interactions. Beyond a quantitative comparison of our results with previous MFT studies, we have characterized the topological nature of the QAH phase by the numerical calculation of a Chern number. In the second part of the work, we have proposed and worked out a scheme for an analog quantum simulation of this model, based on off-resonant laser-coupling of fermionic ground state atoms to electronically excited Rydberg states with strong van-der-Waals interactions. The proposed quantum simulation scheme is implementable with currently available experimental techniques, and it allows one in particular to access the regime of considerably large next-to-nearest neighbor interactions $V_2 \simeq V_1$, which have been identified as one of the essential ingredients for the dynamical emergence of the QAH phase. 

The ideas that we have discussed here for Rydberg-dressed fermionic atoms can also be adapted to the platform of polar molecules, for which a rich toolbox for engineering of effective spin models has been developed \cite{micheli-natphys-2-341,buechler-natphys-3-726,schachenmayer-njp-12-103044}, and which have recently become available in several laboratories\cite{njp-focus-polarmolecules}.

For the quantum simulation of systems with topological properties with cold atoms or molecules in optical lattices, the use of engineered effective interactions to dynamically generate topologically non-trivial phases, as pursued in this work, constitutes a complementary route to the engineering of artificial gauge fields and deserves further exploration. In this context, it will be important to understand the effect of longer-range interaction terms, which are typically to some extent unavoidable in any physical implementation, on the stability of the encountered topological phases. On the other hand, the ample possibilies to engineer a rich variety of interaction potentials raises the question what other types of interaction-driven topological phases could be explored in atomic, molecular and optical quantum simulation setups, and how stable these fascinating novel quantum phases are under realistic couplings to the environment\cite{diehl-nphys-7-971,viyuela-njp-14-033044,arxiv_viyuela-2012}.

\section{Acknowledgments}

A. D. thanks the F.R.S.-FNRS Belgium for financial support and N. Goldman, P. Gaspard, P. de Buyl and G.D. Paparo for support and valuable discussions. We acknowledge support by the Spanish MICINN grant FIS2009-10061, the
CAM research consortium QUITEMAD S2009-ESP-1594, the European Commission
PICC: FP7 2007-2013, Grant No.~249958, and the UCM-BS grant GICC-910758.


\appendix

\section{\label{app:Chern_numbers} Numerical calculation of the Chern numbers}

Here we outline details of the numerical calculation of the Chern numbers based on the algorithm proposed by Fukui \textit{et al.}\cite{bib:fukui} We apply the method to the mean-field Hamiltonian in momentum space (\ref{eq:complete_mean_field_Ham_momentum_space}), with the simplifying physical assumptions discussed in Sect.~\ref{subsec:complete_MF_Ham}, under which the Hamiltonian reduces to
\begin{align}
H_{\rm MF} & = \sum_{\mathbf{k}} 
  \hat{\Psi}^\dagger(\mathbf{k}) \mathcal{H}_{\rm MF}(\mathbf{k}) 
  \hat{\Psi}(\mathbf{k})  \nonumber \\
   & + 3 N_\Lambda \left( V_1 \left( \xi^2 - \overline{n}_\phi \overline{n}_\psi \right) + 2 V_2 | \chi_\phi |^2 \right)  
    \label{eq:complete_mean_field_Ham_momentum_space_app}
\end{align}
with
\begin{align}
\label{eq:Chern_number_Ham_matrix}
\mathcal{H}_{{\rm MF}}(\mathbf{k}) & = 
\begin{pmatrix}
3 V_1 \overline{n}_\psi - 2 V_2 |\chi| f_\mathbf{k}(\pi/2) & - (t + V_1 \xi) A_\mathbf{k}^* \\ 
- (t + V_1 \xi) A_\mathbf{k} & 3 V_1 \overline{n}_\phi -2 V_2 |\chi| f_\mathbf{k}(-\pi/2) 
\end{pmatrix} \nonumber \\
& =  \begin{pmatrix}
3 V_1 \overline{n}_\psi + 2 V_2 |\chi| g_\mathbf{k}(0) & - (t + V_1 \xi) A_\mathbf{k}^* \\ 
- (t + V_1 \xi) A_\mathbf{k} & 3 V_1 \overline{n}_\phi -2 V_2 |\chi| g_\mathbf{k}(0) 
 \end{pmatrix}.
 \end{align}
We work with a discretization of the Brillouin zone according to the grid  $\mathbf{k}=(k_x,k_y)$ with
\begin{align}
k_x&=-\frac{2\pi}{3}+n_x\,s, \nonumber \\
k_y&=-\frac{2\pi}{3\sqrt{3}}+n_y\,s,
\end{align}
where the integer-valued indices $n_x$, $n_y$ run as
\begin{align}
0&<n_x \leqslant \lfloor 4\pi/(3s) \rfloor, \nonumber \\
0&<n_y\leqslant \lfloor 2\pi/(\sqrt{3}s) \rfloor,
\end{align}
such that the mesh of $\mathbf{k}$-vectors covers the complete Brillouin zone. The parameter $s$ determines the degree of coarsening of the mesh, and is fixed to $s=0.1$ in our calculations. See also Appendix~\ref{app:brillouin_zone} for details on this and other equivalent parametrizations of the Brillouin zone.

On this discrete set of points $\{\mathbf{k}\}$ we define the U(1) quantum link variables 
\begin{equation}
\label{eq:quantum_link_variable}
U_\mu(\mathbf{k}) \equiv \langle u(\mathbf{k}) | u(\mathbf{k} +\hat{\mu} )\rangle/N_\mu.
\end{equation}
Here, $\hat{\mu}$ is a vector pointing to the next mesh point in the direction $k_x$ or $k_y$, $N_\mu$ a normalization constant of the hermitian product and $|u(\mathbf{k})\rangle$ is the normalized eigenvector corresponding to the lower-band eigenvalue $E_{-}(\mathbf{k})$ of matrix (\ref{eq:Chern_number_Ham_matrix}),
\begin{equation}
E_{-}(\mathbf{k}) = 3V_1 n - \left[ \left( 3V_1 \rho -2 V_2 |\chi| g_\mathbf{k}(0) \right)^2 + t'^2 |A_\mathbf{k}|^2 \right]^{1/2},
\end{equation}
which can be readily obtained in analytical form. From the quantum link variables we then calculate the lattice field strength
\begin{equation}
\label{eq:lattice_field_strength}
 \tilde{F}_{xy}(\mathbf{k}) \equiv \ln U_x (\mathbf{k}) U_y (\mathbf{k}+\mathbf{1}_{k_x}) U_x(\mathbf{k} + \mathbf{1}_{k_y})^{-1} U_2(\mathbf{k})^{-1}.
\end{equation}

As emphasized in Ref.\cite{bib:fukui}, it is important and always possible to choose a grid where for all $\mathbf{k}$-vectors the eigenvectors of Hamiltonian (\ref{eq:complete_mean_field_Ham_momentum_space_app}) are well-defined. Inspection of Eq.~(\ref{eq:Chern_number_Ham_matrix}) shows that this can in the present case be guaranteed if a mesh is chosen where the points of the Dirac cones are excluded (see Appendix~\ref{app:brillouin_zone} for explicit expressions for their locations.)

By evaluation of expressions (\ref{eq:quantum_link_variable}) and (\ref{eq:lattice_field_strength}) over the discrete set of points parametrizing the Brillouin zone, it is then straightforward to compute the Chern number for the system at half-filling ($n=1/2$), with the lower (upper) band completely filled (empty), according to the sum over the discretized Brillouin zone
\begin{equation}
\tilde{c}_{1} \equiv \frac{1}{2\pi i} \sum_{\text{grid} \,\{ \mathbf{k}\}} \tilde{F}_{xy}(\mathbf{k}).
\end{equation}

Finally, we remark that the numerical algorithm is very efficient and sensitive in the sense that even for non-zero $\chi$ values on the order of $10^{-10}$, it yields a Chern number $c_{1} = 1$. As the precise value of $\chi$ depends on the finite, numerical accuracy of determining the mean values for $\chi$ from either minimizing the free energy (\ref{eq:total_free_energy_simplified}) or solving the self-consistency equations (\ref{eq:xi_self_consistent})-(\ref{eq:chi_self_consistent}), we introduce a small, but finite cut-off below which we set non-zero $\chi$-solutions to zero, to avoid spurious detection of non-zero Chern number values, originating from finite numerical precision.

\section{\label{app:brillouin_zone} Details on the Brillouin zone and domain of integration}

\begin{figure}[t]
	\includegraphics[width=\columnwidth]{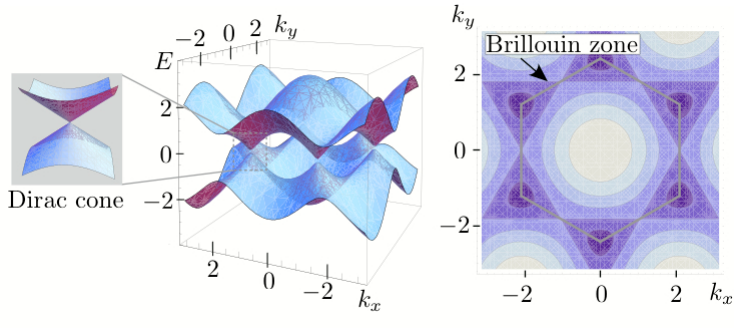}\hspace{0.8cm}
	\caption{Left: Energy spectrum over the first Brillouin zone for vanishing staggering potential $\beta=0$ and $t=1$. Right: Color-coded contour plot of the energy spectrum.}
	\label{fig:diraccone}
\end{figure} 

In this appendix we provide explicit expressions and point out useful symmetry properties of various, equivalent parametrizations of the Brillouin zone of the hexagonal lattice structure (cf.~Fig.~\ref{fig:honeycomb}). In particular, we provide the parametrization which is used for the numerical calculation of the Chern number based on a discretized Brillouin zone. 
\begin{figure}[t]
	\includegraphics[width=\columnwidth]{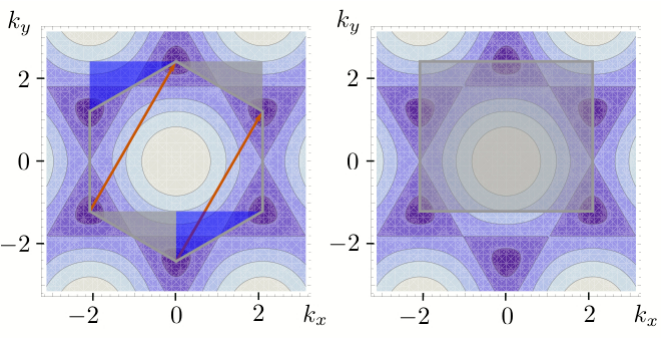}\hspace{0.8cm}
	\caption{The two blue- and gray-shaded triangles at the bottom of the hexagonal-shaped first Brillouin zone can be translated to the top of the hexagon to obtain an equivalent, rectangular-shaped area of integration.}
	\label{fig:bzv3}
\end{figure} 
\begin{figure}
	\includegraphics[width=0.4\columnwidth]{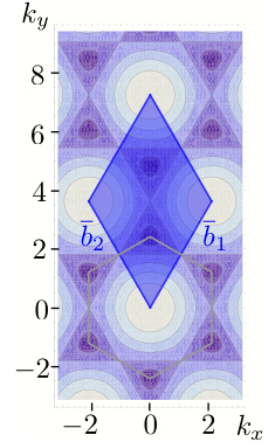}\hspace{0.8cm}
	\caption{Comparison of the Brillouin zone defined as the Wigner-Seitz unit cell of the reciprocal lattice (gray, hexagonal area), and by an alternative, though equivalent definition as the area of the parallelogram spanned by the vectors $b_1$ and $b_2$ (blue-shaded area).}
	\label{fig:bzv2}
\end{figure} 

For the basis vectors of the real-space honeycomb lattice $\mathbf{a_1}=(3/2,\sqrt{3}/2)$ and $\mathbf{a_2}=(-3/2,\sqrt{3}/2)$ (lattice spacing $a$ set to one), the basis vectors of the reciprocal lattice are defined by $\mathbf{a_i}.\mathbf{b_j}=2\pi \delta_{\text{ij}}$ and given by 
$\mathbf{b_1}=(2\pi/3,2\pi/\sqrt{3})$ and $\mathbf{b_2}=(-2\pi/3,2\pi/\sqrt{3})$. The first Brillouin zone is a hexagon, for which the positions of the vertices are:
\begin{align}
& (0,\frac{4\pi}{3\sqrt{3}}),\,\, (\frac{2\pi}{3},\frac{2\pi}{3\sqrt{3}}),\,\, (\frac{2\pi}{3},-\frac{2\pi}{3\sqrt{3}}), \nonumber \\
& (0,-\frac{4\pi}{3\sqrt{3}}),\,\, (-\frac{2\pi}{3},-\frac{2\pi}{3\sqrt{3}}),\,\, (-\frac{2\pi}{3},\frac{2\pi}{3\sqrt{3}}).
\end{align}
The volume of the Wigner-Seitz cell of the real lattice, i.e.~the area corresponding to one two-site basis cell, is $3\sqrt{3}/2$; thus the 
area of the domain of integration over the first Brillouin zone is $L^2=(2\pi)^2 /( 3\sqrt{3}/2) = 8\pi^2/(3\sqrt{3})$. Fig.~\ref{fig:diraccone} shows the two-band energy spectrum in the first Brillouin zone for vanishing staggering potential $\beta = 0$. 

Sometimes, in particular for comparison with results of related work in the literature\cite{Raghu-prl-100-156401,Weeks-prb-81-085105}, it is convenient to move from the discrete sum over momenta to the integral over the Brillouin zone. Noting that
\begin{equation}
\sum_{\mathbf{k} \in \text{BZ}} 1 = N_\Lambda, \qquad \mathrm{and} \qquad \int_{\text{BZ}} \mathrm{d}^2k = L^2,
\end{equation}
this change can be realized according to
\begin{equation}
\sum_{\mathbf{k} \in \text{BZ}} \ldots \rightarrow \frac{N_\Lambda}{L^2} \int_{\text{BZ}} \mathrm{d}^2k \ldots
\end{equation}

Furthermore, by making use of the translational invariance of the Brillouin zone, one can change from the hexagonal-shaped Brillouin zone to a domain of integration over a rectangular-shaped integration area, as indicated in Fig.~\ref{fig:bzv3}. 

Whereas the Brillouin zone, as shown in the left panel of Fig.~\ref{fig:bzv2}, is defined in the standard way as the Wigner Seitz unit cell in the reciprocal lattice, it can alternatively also be defined as the parallelogram spanned by the vectors $\mathbf{b}_1$ and $\mathbf{b}_2$ - as indicated by the blue-shaded area in Fig.~\ref{fig:bzv2}). In the present case of the hexagonal lattice the area is even more symmetric (a rhombus), since $\| \mathbf{b}_1 \| = \| \mathbf{b}_2 \|$. 

Whereas the domains of integration of the rectangular-shaped and the parallelogram-like Brillouin zones are more suitable for numerical calculations, symmetry properties are most explicit in the standard definition (hexagonal-shaped Brillouin zone). 



%

\end{document}